\documentclass[11pt]{article}
\usepackage{subfigure}
\usepackage{epsf}
\usepackage{psfrag}
\usepackage{cite}
\usepackage{amssymb}
\usepackage{amsmath}
\usepackage{mathrsfs}
\usepackage{graphicx}
\usepackage{latexsym}
\usepackage{array}

\newtheorem{theorem}{Theorem}

\newtheorem{corollary}{Corollary}
\newtheorem{lemma}{Lemma}
\newcommand{\eqdef}{\stackrel{\triangle}{=}}
\newcommand{\bc}{{\bf c}}

\newcommand{\mC}{{\mathcal C}}
\newcommand{\mQ}{{\mathcal Q}}
\newcommand{\bmQ}{{\bar {\mathcal Q}}}
\newcommand{\hmQ}{{\hat {\mathcal Q}}}

\newcommand{\rs}{\hspace*{-0.1in} }
\newcommand{\aline}[1]{\rs&{#1}&\rs}
\newcommand{\GF}{\text{GF}}

\newcommand{\hsp}{\hspace{0.2in}}

\begin{document}
\bibliographystyle{unsrt}
\renewcommand{\baselinestretch}{1.35}\normalsize
\begin{titlepage}

\thispagestyle{empty}

\title{\bf New List Decoding Algorithms for Reed-Solomon and BCH Codes}

\author{Yingquan Wu \\[0.1in]
Link\_A\_Media Devices Corp. \\[0.1in]
}


\end{titlepage}
\maketitle

\begin{abstract}
In this paper we devise a rational curve fitting algorithm and apply it to the list decoding of Reed-Solomon and BCH codes. 
The resulting list decoding algorithms exhibit the following significant properties.
\begin{itemize}
\item The algorithm achieves the limit of list error correction capability (LECC) $n(1-\sqrt{1-D})$ for a 
(generalized) $(n, k, d=n-k+1)$ Reed-Solomon code, which matches the Johnson bound,
 where $D\eqdef \frac{d}{n}$ denotes the normalized minimum distance. 
The algorithmic complexity is $O\left(n^{6}(1-\sqrt{1-D})^8\right)$.
In comparison with the Guruswami-Sudan algorithm, which exhibits the same LECC,
the proposed requires a multiplicity (which dictates the algorithmic complexity) significantly smaller than 
that of the Guruswami-Sudan algorithm in achieving a given LECC, except for codes with code-rate below 0.15. 
In particular, for medium-to-high rate codes, the proposed algorithm reduces the multiplicity by orders of magnitude.
Moreover, 
for any $\epsilon>0$, the intermediate LECC $t=\lfloor\epsilon\cdot \frac{d}{2} + (1-\epsilon) \cdot (n-\sqrt{n(n-d)})\rfloor$ 
can be achieved by the proposed algorithm with multiplicity $m=\lfloor\frac{1}{\epsilon}\rfloor$. 
Its list size is shown to be upper bounded by a constant with respect to a fixed normalized minimum distance $D$, 
rendering the algorithmic complexity quadratic in nature, $O(n^2)$.

\item By utilizing the unique properties of the Berlekamp algorithm, the algorithm achieves the LECC limit
 $\frac{n}{2}(1-\sqrt{1-2D})$ for a narrow-sense $(n, k, d)$ binary BCH code, which matches the Johnson bound for binary codes. 
The algorithmic complexity is $O\left(n^{6}(1-\sqrt{1-2D})^8\right)$.
Moreover, for any $\epsilon>0$, the intermediate LECC $t=\lfloor\epsilon\cdot \frac{d}{2} + (1-\epsilon) \cdot \frac{n-\sqrt{n(n-2d)}}{2}\rfloor$ 
can be achieved by the proposed algorithm with multiplicity $m=\lfloor\frac{1}{2\epsilon}\rfloor$. 
Its list size is shown to be upper bounded by a constant, 
rendering the algorithmic complexity quadratic in nature, $O(n^2)$.

\end{itemize}

\end{abstract}


\newpage

\section{\sc Introduction}

Reed-Solomon codes are the most commonly used error correction codes in practice. 
Their widespead applications include magnetic and optical data storage, wireline and wireless communications, and satellite communications.
A Reed-Solomon code $(n, k)$ over a finite field $\GF(q)$ satisfies $n<q$ and achieves the maximally separable distance, i.e., $d=n-k+1$. 
Its algebraic decoding has been extensively explored but remains a challenging research topic. 

It is well-known that efficient algorithms exist to decode up to half the minimum distance with complexity $O(dn)$, namely,
the Berlekamp-Massey algorithm \cite{Berlekamp, Massey} and the Euclidean algorithm \cite{Sugiyama}, 
which utilize the frequency spectrum property,  and the Berlekamp-Welch algorithm, which utilizes the polynomial characteristics \cite{Welch}.
Koetter \cite{Koetter} devised a one-pass algorithm, building on the Berlekamp-Massey algorithm, 
to implement the generalized minimum distance (GMD) decoding, which otherwise requires $\lfloor \frac{d+1}{2} \rfloor$ rounds.
Berlekamp \cite{Berlekamp2} devised a one-pass algorithm, building on the Berlekamp-Welch algorithm, to implement $d+1$ GMD decoding.
Kamiya \cite{Kamiya} presented one-pass GMD decoding algorithms and a one-pass Chase decoding algorithm for BCH codes 
utilizing the Berlekamp-Welch algorithm.

Sudan \cite{Sudan} discovered a polynomial-time algorithm, building on the Berlekamp-Welch algorithm, 
for (list) decoding Reed-Solomon codes beyond the classical correction capability $\lfloor\frac{d-1}{2}\rfloor$,  
however, the algorithm is effective only when the code rate $\frac{k}{n}<\frac{1}{3}$. 
Schmidt, Sidorenko, and Bossert \cite{Schmidt} virtually extend one codeword to a sequence of interleaved codewords 
which yields a multiple-sequence linear shift register synthesis, and exploit
the generalized Berlekamp-Massey algorithm, whose complexity is quadratic in nature, to correct errors beyond half the minimum distance. 
The algorithm succeeds if there is a unique solution within a certain capability, 
which is larger than the conventional error correction capability when the code rate is below $\frac{1}{3}$.
Its error correction capability and rate threshold largely coincide with those of the Sudan algorithm in \cite{Sudan},
whereas its algorithmic complexity is much lower than the Sudan algorithm. 
Guruswami and Sudan \cite{Guru-Sudan} devised an improved version of \cite{Sudan}, 
which is capable of decoding beyond half the minimum distance over all rates. 
More specifically, the algorithm lists all codewords up to distance $n(1-\sqrt{1-D})$ (where $D$ denotes 
the normalized minimum distance $D\eqdef \frac{d}{n}$)  from the received word,
while the algorithmic complexity is polynomial in nature. Its performance matches the Johnson bound \cite{Johnson},
which gives a general lower bound on the number of errors one can correct using small lists in any code, 
as a function of the normalized minimum distance $D$. 
McEliece \cite{McEliece} characterized the average list size of the Guruswami-Sudan algorithm and 
showed that the list most likely contains only one codeword.
Guruswami and Rudra \cite{Guru} showed the optimality of the list error correction capability (LECC) 
$n(1-\sqrt{1-D})$ in the sense that the number of codewords lying
slightly beyond the boundary can be superpolynomially large in code length $n$.
Koetter and Vardy \cite{Koetter-Vardy} showed a natural way to translate the soft-decision reliability information provided 
by the channel into the multiplicity matrix which is directly involved in the Guruswami-Sudan algorithm. 
The resulting algorithm outperforms the Guruswami-Sudan algorithm.

In essence, the Guruswami-Sudan algorithm is a polynomial curve-fitting algorithm that determines all polynomials 
which passes through at least $\sqrt{n(n-d)}$ points out of $n$ distinct points.
Specifically, when given $n$ distinct points $\{ (x_i, y_i) \}_{i=0}^{n-1}$, where $[y_0, y_1, \ldots, y_{n-1}]$ 
denotes a received word, a $(1, n-d)$-weighted degree bivariate polynomial 
$\mQ(x, y)$ is constructed to pass through all $n$ points, each with appropriate multiplicity, then $\mQ(x, y)$ contains
 all desired polynomials $p(x)$ as its factors in the form of $y-p(x)$.
Finally all desired polynomials $p(x)$ are factorized iteratively \cite{Guru-Sudan}. 
The interpolation process can be expedited by utilizing the updating algorithm in \cite{Koetter} with quadratic complexity, 
whereas straightforward implementation using Gaussian elimination requires cubic complexity.

\begin{figure}[t] 
\centering
\includegraphics[width=6.4in]{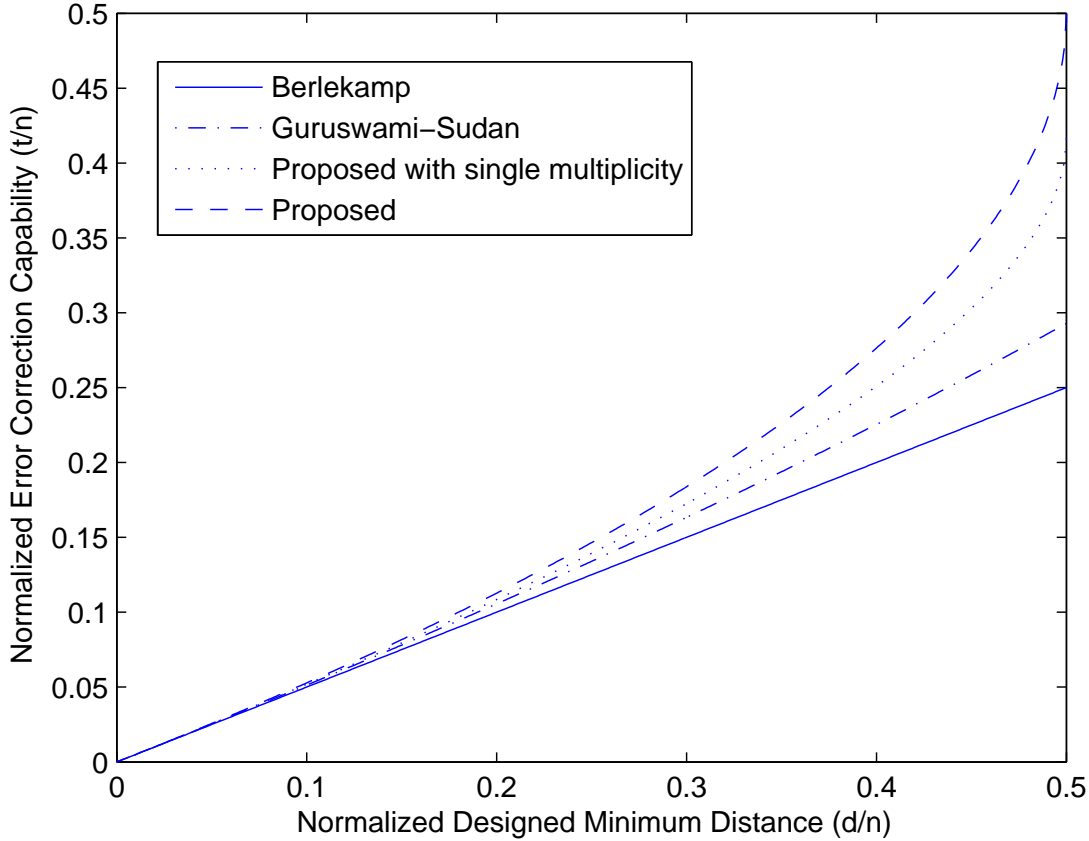}
\caption{Normalized (list) error correction capability as a function of the normalized designed minimum distance for binary BCH codes. 
\label{FIG-BCH-bound}}
\end{figure}

In this paper, we formulate the list decoding of Reed-Solomon codes as a rational curve-fitting problem utilizing the polynomials constructed 
by the Berlekamp-Massey algorithm. 
Specifically, let $\Lambda(x)$ and $B(x)$ be the error locator and correction polynomials, respectively, as obtained from the Berlekamp-Massey algorithm.
List decoding then finds pairs of polynomials $b(x)$ and $\lambda(x)$, 
such that each pair leads to a valid candidate error locator polynomial
$\Lambda'(x)\eqdef \lambda(x) \cdot \Lambda(x) + b(x) \cdot xB(x)$, i.e., 
all the roots of $\Lambda'(x)$ are distinct and belong to the pre-defined space.
We reduce this problem to a rational curve-fitting problem and subsequently present a novel polynomial algorithm 
comprised of rational interpolation and rational factorization.
Using the up-to-date most efficient implementation algorithms \cite{Koetter, Ma},
the proposed algorithm exhibits the complexity $O\left(n^{2}(\sqrt{n}-\sqrt{k})^8\right)$.
The proposed list decoding algorithm exhibits the same LECC $n-\sqrt{n(n-d)}$ as the Guruswami-Sudan algorithm. 
However, the proposed algorithm requires much lower multiplicity, which dictates the algorithmic complexity, 
for almost the entire range of code rates. 
In particular, the proposed algorithm reduces the multiplicity by orders of magnitude for medium-to-high rate codes.
Finally, the proposed algorithm utilizes the end results of the Berlekamp-Massey algorithm, 
in contrast to the decoding algorithms in \cite{Roth, Schmidt}, 
which directly incorporate syndromes and achieve performance gains over conventional hard-decision decoding only 
when the code rate $\frac{k}{n}<\frac{1}{3}$.  
By observing that even iterations in the Berlekamp algorithm are automatically satisfied 
in decoding  binary BCH codes, we present a modified version of the proposed algorithm 
that exhibits the LECC $\frac{n}{2}(1-\sqrt{1-2D})$, 
which matches the Johnson bound for binary codes \cite{Johnson}. 
A comparison to existing state-of-the-art LECCs is depicted in Figure~\ref{FIG-BCH-bound}.

We also reveal a fundamental property of Reed-Solomon and binary BCH codes, namely,
 that there exist at most a constant number of codewords, regardless of code length $n$,
 with respect to an LECC under arbitrary small fraction away from the Johnson bound. 
 Furthermore, we show that the corresponding Johnson bound can be arbitrarily approximated by the derivative algorithms 
with quadratic complexity, for both Reed-Solomon and binary BCH codes.

The remainder of the paper is organized as follows. Section II.A briefly introduces  
the Berlekamp-Massey algorithm for decoding Reed-Solomon codes,  and then extends one iteration to 
correct up to $\lfloor\frac{d}{2}\rfloor$ errors with negligible additional complexity.
Section II.B presents a re-formulated Berlekamp algorithm for decoding binary BCH codes, and then extends one iteration to 
correct up to $\lfloor\frac{d+1}{2}\rfloor$ errors with negligible additional complexity.
The proposed list decoding algorithm for Reed-Solomon codes is detailed in Section III.
The list decoding problem is formulated in Part A, the rational interpolation process is then described in Part B,
 followed by the rational factorization in Part C. 
The algorithmic description and performance assertion are presented in Part D and 
the computational complexity is characterized in Part E. Finally Part F shows that 
the LECC limit can arbitrarily approximated with derivative algorithms with constant multiplicities 
which exhibit only quadratic complexity. 
Section IV presents an improved algorithm for 
decoding  binary BCH codes. The paper is concluded with pertinent remarks in Section V.


\section{\sc Algebraic Hard-Decision Decoding of Reed-Solomon and BCH Codes}

\subsection{Berlekamp-Massey Algorithm and its One-Step Extension for Decoding Reed-Solomon Codes}

For a (possibly shortened) Reed-Solomon $\mC(n, k)$ code over $\GF(q)$, a $k$-symbol  ${\bf D}\eqdef [D_{k-1}$, $D_{k-2}$, \ldots,  $D_1$, 
$D_0]$ is encoded to an $n$-symbol codeword ${\bf C} \eqdef [C_{n-1}$, $C_{n-2}$, \ldots,  $C_1$,  $C_0]$, or more conveniently, a dataword polynomial 
$D(x)=D_{k-1} x^{k-1} + D_{k-2} x^{k-2} + \ldots +D_1 x^1 + D_0$ is encoded to a codeword polynomial $C(x)=C_{n-1} x^{n-1} +C_{n-2} x^{n-2} + \ldots +
 C_1 x +C_0 $,
by means of a generator polynomial 
$$G(x)\eqdef \prod_{i=0}^{n-k-1} (x-\alpha^{m_0+i})$$
where $\alpha$ is a primitive element of $\GF(q)$ and $m_0$ is an arbitrary integer 
(in this presentation we do not distinguish between a vector ${\bf A}=[A_0, A_1, A_2, \ldots, A_l]$ and its polynomial representation 
$A(x)=A_0+A_1x +A_2x^2 +\ldots A_l x^l$). 
A polynomial of degree less than $n$ is 
a codeword polynomial if and only if it is a multiple of the generator polynomial $G(x)$. 
As can be readily seen, a codeword polynomial $C(x)$ satisfies
$$C(\alpha^{i})=0, \hspace{0.15in} i=m_0, m_0+1, m_0+2,\ldots, m_0+n-k-1. $$
The minimum Hamming distance of the code is $d=n-k+1$, an attribute known as {\em maximally-distance-separable} (cf. \cite{Blahut}).

Let $C(x)$ denote the transmitted codeword polynomial and  $R(x)$ the received word polynomial. 
The decoding objective is to determine the error polynomial $E(x)$ such that $C(x)=R(x)-E(x)$. 

In the following we introduce the Berlekamp-Massey algorithm, which provides a foundation for our list decoding algorithms. 
It begins with the task of error correction 
by computing syndrome values
$$S_i=R(\alpha^{i+m_0})=C(\alpha^{i+m_0})+E(\alpha^{i+m_0})=E(\alpha^{i+m_0}), \hspace{0.15in} i=0, 1, 2, \ldots, n-k-1.$$
If all $n-k$ syndrome values are zero, then $R(x)$ is a codeword polynomial and thus is presumed that $C(x)=R(x)$, i.e., no errors have occurred. 
Otherwise, let $e$ denote the (unknown) number of errors, $X_i\in \{\alpha^{-i}\}_{i=0}^{n-1}$, $i=1, 2, \ldots, e$, denote 
the  error locations,  and $Y_i\in \GF(q)$, $i=1, 2, \ldots, e$, denote the corresponding error magnitudes.

Define the {\em syndrome} polynomial
\begin{equation}
S(x)\eqdef S_0 +S_1 x +S_2 x^2 +\ldots +S_{n-k-1} x^{n-k-1},
\end{equation}
 the {\em error locator} polynomial 
\begin{equation}
\Lambda(x)\eqdef \prod_{i=1}^e (1-X_ix)=1+\Lambda_1 x + \Lambda_2 x^2 +\ldots +\Lambda_e x^e,  \label{def-true-Lambda}
\end{equation}
and the {\em error evaluator} polynomial 
\begin{equation}
\Omega(x)\eqdef \sum_{i=1}^e Y_i X_i^{m_0} \prod_{j=1, j\ne i}^e (1-X_jx)=\Omega_0+ \Omega_1 x +\Omega_2 x^2+\ldots +\Omega_{e-1} x^{e-1}. \label{def-Omega}
\end{equation}
The three polynomials satisfy the following {\em key equation} (cf. \cite{Blahut})
\begin{equation}
\Omega(x)= \Lambda(x)S(x)  \hspace{0.2in} \pmod{x^{n-k}}.  \label{key-equ}
\end{equation}

The Berlekamp-Massey algorithm  can be used to solve the above key equation, given 
that the number of errors $e$ does not exceed the error-correction capability $\lfloor \frac{n-k}{2}\rfloor$ (cf. \cite{Berlekamp, Blahut}).
Below we slightly re-formulate the Berlekamp-Massey algorithm given in \cite{Blahut},
so as to facilitate the characterizations afterwards:

\vspace{0.1in}
{\underline {\bf Berlekamp-Massey Algorithm}}
{\small \begin{itemize}
\item Input:\hsp ${\bf S}=[S_0, \;\; S_1,  \; \;S_2,\; \;\ldots,\;\; S_{n-k-1}]$
\item Initialization:\hsp $\Lambda^{(0)}(x)=1$, $B^{(0)}(x)=1$, and $L^{(0)}_\Lambda=0$, $L^{(0)}_B=0$
\item For $r=0$, 1, 2, \ldots, $n-k-1$, \hsp do:
\begin{itemize}
\item Compute $\Delta^{(r+1)}=\sum_{i=0}^{L^{(r)}_\Lambda} \Lambda^{(r)}_i \cdot S_{r-i}$
\item Compute $\Lambda^{(r+1)}(x)=\Lambda^{(r)}(x) -\Delta^{(r+1)} \cdot x B^{(r)}(x)$
\item If $\Delta^{(r+1)} \ne 0$ and $2L^{(r)}_\Lambda\leq r$, \hsp then
\begin{itemize}
\item Set $B^{(r+1)}(x)\gets (\Delta^{(r+1)})^{-1} \cdot\Lambda^{(r)}(x)$ 
\item Set $L^{(r+1)}_\Lambda\gets L^{(r)}_B+1$, \ $L^{(r+1)}_B\gets L^{(r)}_\Lambda$
\end{itemize}
\item Else 
\begin{itemize}
\item Set $B^{(r+1)}(x) \gets x B^{(r)}(x)$
\item Set $L^{(r+1)}_B\gets L^{(r)}_B+1$, \ $L^{(r+1)}_\Lambda\gets L^{(r)}_\Lambda$
\end{itemize}
\item[] endif
\end{itemize}
\item[] endfor
\item Output:\hsp $\Lambda(x)$, \ $B(x)$, \ $L_\Lambda$, \ $L_B$
\end{itemize}
}

Note that in the above description, 
we used superscript ``$^{(r)}$" to stand for the $r$-th iteration and subscript ``$_i$'' the $i$-th coefficient. 
$L_\Lambda$ and $L_B$ denote the length of linear feedback shift register (LFSR) described by $\Lambda(x)$ and $B(x)$, respectively. 
An LFSR of length $L$, $a_0=1$,
$a_1$, $a_2$, \ldots, $a_L$, is called to {\em generate} the sequence $s_0$, $s_1$, $s_2$, \ldots, $s_{r}$ if 
\begin{equation}
\sum_{j=0}^L s_{i-j} a_j=0,   \hsp i=L, L+1, \ldots, r. \label{def-L}
\end{equation}

The essence of the Berlekamp-Massey algorithm is to determine a minimum-length  LFSR 
that {\em generates} the syndrome sequence $S_0$, $S_1$, $S_2$, \ldots, $S_{n-k-1}$ \cite{Berlekamp, Massey}, 
It is worth mentioning that there may exist multiple minimum-length LFSRs that generate the sequence $S_0, S_1, \ldots, S_{n-k-1}$, 
and $\Lambda(x)$ obtained from the Berlekamp-Massey algorithm is one of them when non-unique.
The error locator polynomial $\Lambda(x)$ and the correction polynomial $B(x)$ are characterized by the following lemma.
\begin{lemma}   \label{LEM-BMA-character}
Let $\Lambda(x)$ be the error locator polynomial and $B(x)$ be the correction polynomial, computed by the Berlekamp-Massey algorithm.\\
$(i)$. $L_\Lambda$ and $L_B$, the length of LFSR described by $\Lambda(x)$ and $B(x)$ respectively,  satisfy 
\begin{equation}
L_\Lambda + L_B=d-1.
\end{equation}
$(ii)$. The degrees of $\Lambda(x)$ and $B(x)$ satisfy   
\begin{equation}
\deg(\Lambda(x)) \leq L_\Lambda, \hspace{0.3in} \deg(B(x)) \leq L_B.
\end{equation}
When $\Lambda(x)$ is the true error locator polynomial as defined in \eqref{def-true-Lambda}, $\deg(\Lambda(x))=L_\Lambda$.\\
$(iii)$. The polynomials  $\Lambda(x)$ and $B(x)$ are coprime, i.e., the two do not share a common factor. 
\end{lemma}
{\em Proof:} 
$(i)$. We show the following more general result
\begin{equation}
L^{(r)}_{B} + L^{(r)}_{\Lambda}=r.  \label{general-L-sum}
\end{equation}
It follows that in each iteration either $L^{(r+1)}_\Lambda\gets L^{(r)}_B+1$, $L^{(r+1)}_B\gets L^{(r)}_\Lambda$,  
or  $L^{(r+1)}_\Lambda\gets L^{(r)}_\Lambda$, $L^{(r+1)}_B\gets L^{(r)}_B+1$, thus their sum increases by 1 in either case. \\
$(ii)$. We show the first part by induction. When $i=0$, we have $L^{(0)}_{\Lambda}=\deg(\Lambda^{(0)}(x))=0$ and $L^{(0)}_{B}=\deg(B^{(0)}(x))=0$. 
Assume that $\deg(\Lambda^{(r)}(x)) \leq L^{(r)}_\Lambda  $ and $\deg(B^{(r)}(x))\leq L^{(r)}_B$ hold for $i=r$. 
In the case of $\Delta^{(r+1)} \ne 0$ and $2L^{(r)}_{\Lambda} \leq r$, we have the following iteration
\begin{itemize}
\item $\Lambda^{(r+1)}(x)=\Lambda^{(r)}(x) -\Delta^{(r+1)} \cdot x B^{(r)}(x)$
\item $B^{(r+1)}(x)\gets (\Delta^{(r+1)})^{-1} \cdot\Lambda^{(r)}(x)$ 
\item $L^{(r+1)}_{\Lambda}  \gets L^{(r)}_{B}+1$, $L^{(r+1)}_{B}\gets L^{(r)}_{\Lambda}$. 
\end{itemize}
\eqref{general-L-sum}, in conjunction with the condition $2L^{(r)}_{\Lambda} \leq r$, results in
$$L^{(r)}_{B}=r- L^{(r)}_{\Lambda} \geq  L^{(r)}_{\Lambda}.$$
Therefore, we obtain
$$\deg(\Lambda^{(r+1)}(x)) =\max\{ \deg(\Lambda^{(r)}(x)), \;  \deg(B^{(r)}(x))+1\} 
\leq \max\{ L^{(r)}_{\Lambda},\; L^{(r)}_{B}+1\} = L^{(r)}_{B}+1=L^{(r+1)}_{\Lambda}.$$
$$\deg(B^{(r+1)}(x)) = \deg(\Lambda^{(r)}(x)) \leq  L^{(r)}_{\Lambda} = L^{(r+1)}_{B}.$$
When $\Delta^{(r+1)}=0$, we have the following update
\begin{itemize}
\item $B^{(r+1)}(x) \gets x B^{(r)}(x)$, \ \ $L^{(r+1)}_B\gets L^{(r)}_B+1$.
\end{itemize}
The conclusion naturally holds. Finally, when  $\Delta^{(r+1)} \ne 0$ and $2L^{(r)}_{\Lambda} > r$, the algorithmic updates follow
\begin{itemize}
\item $\Lambda^{(r+1)}(x)=\Lambda^{(r)}(x) -\Delta^{(r+1)} \cdot x B^{(r)}(x)$
\item $B^{(r+1)}(x) \gets x B^{(r)}(x)$, \ \  $L^{(r+1)}_B\gets L^{(r)}_B+1$.
\end{itemize}
\eqref{general-L-sum}, in conjunction with the condition $2L^{(r)}_{\Lambda} > r$, results in
$$L^{(r)}_{B}=r- L^{(r)}_{\Lambda} <  L^{(r)}_{\Lambda}.$$
Therefore, we obtain 
$$\deg(\Lambda^{(r+1)}(x)) =\max\{ \deg(\Lambda^{(r)}(x)), \;  \deg(B^{(r)}(x))+1\} 
\leq \max\{ L^{(r)}_{\Lambda},\; L^{(r)}_{B}+1\} \leq L^{(r)}_{\Lambda}=L^{(r+1)}_{\Lambda}$$
$$\deg(B^{(r+1)}(x)) =\deg(B^{(r)}(x))+1 \leq L^{(r)}_{B}+1 = L^{(r+1)}_{B}.$$
We thus have justified the first part of $(ii)$.  
The second part naturally follows \eqref{def-Omega}, \eqref{key-equ} and the definition of generating a sequence in \eqref{def-L}. 

Part $(iii)$ can be shown through contradiction (cf. \cite{Koetter}). Herein we give an inductive proof.
Evidently, when $i=0$,  $\Lambda^{(0)}(x)=1$ and $B^{(0)}(x)=1$ are coprime. Assume that $\Lambda^{(r)}(x)$ and $B^{(r)}(x)$ are coprime for $i=r$.
For $i=r+1$, if $\Delta^{(r+1)} \ne 0$ and $2L^{(r)}_{\Lambda} \leq r$, then the iteration, 
$\Lambda^{(r+1)}(x)=\Lambda^{(r)}(x) -\Delta^{(r+1)} \cdot x B^{(r)}(x)$ and 
$B^{(r+1)}(x)\gets (\Delta^{(r+1)})^{-1} \cdot\Lambda^{(r)}(x)$, clearly indicates that $\Lambda^{(r+1)}(x)$ and $B^{(r+1)}(x)$ are coprime, 
conditioned on that $\Lambda^{(r)}(x)$ and $B^{(r)}(x)$ are coprime; so is the alternative iteration, 
$\Lambda^{(r+1)}(x)=\Lambda^{(r)}(x) -\Delta^{(r+1)} \cdot x B^{(r)}(x)$ 
and $B^{(r+1)}(x) \gets x B^{(r)}(x)$.   We conclude that $\Lambda(x)$ and $B(x)$ are coprime.      \hfill $\Box\Box$

Note the initial cause of $\deg(\Lambda(x))<L_\Lambda$ is due to the special condition \cite{Berlekamp}
$$L^{(r)}_\Lambda= L^{(r)}_B+1 \text{ \ and \ } \Lambda^{(r)}_{l}=\Delta^{(r+1)} B^{(r)}_{l-1}, \text{ where } l=L^{(r)}_\Lambda.$$

Let $n-k$ be an odd number. Then, the number of errors up to 
\begin{equation}
t_0\eqdef \frac{n-k+1}{2}=\frac{d}{2}  \label{def-t0}
\end{equation}
can be corrected by the following simple list decoding algorithm

{\underline {\bf One-Step-Ahead Berlekamp-Massey Algorithm}}
\begin{enumerate}
\item If $L_\Lambda > t_0$, then declare a decoding failure. 
\item If $L_\Lambda < t_0$, then determine all distinct roots in $\{\alpha^{-i}\}_{i=0}^{n-1}$. 
If the number of (distinct) roots is equal to $L_\Lambda$, then apply Forney's formula and return 
the unique codeword, otherwise declare a decoding failure (which is identical to the normal Berlekamp-Massey algorithm).
\item Evaluate $\Delta_i=\frac{\Lambda(\alpha^{-i})}{\alpha^{-i}B(\alpha^{-i})}$, $i=0, 1, 2, \ldots, n-1$.
\item Group the index sets $\{i_1, i_2, \ldots, i_{t_0} \}$ such that $\Delta_i$'s are identical 
(each set corresponds to the roots of a valid error locator polynomial).
\item Apply Forney's formula to compute error magnitudes  with respect to each index set, each resulting in a candidate codeword. 
\end{enumerate}

{\em Proof of correctness: } We note that $\Lambda(x)$ and $B(x)$ are obtained at the $(n-k)$-th iteration of the Berlekamp-Massey algorithm. 
Following the nature of the Berlekamp-Massey algorithm, 
the additional syndrome $S_{n-k}$ determines all valid error locator polynomial $\Lambda^*(x)$ of degree up to $t_0$.
More specifically,  a valid error locator polynomial $\Lambda^*(x)$ of degree up to $t_0$ satisfies the form
\begin{equation}
\Lambda^*(x)=\Lambda(x)-\Delta(S_{n-k})\cdot xB(x) \label{one-step}
\end{equation}
where the discrepancy $\Delta(S_{n-k})$ is a linear function of $S_{n-k}$, 
$$\Delta(S_{n-k})=\sum_{i=0}^{L_\Lambda} S_{n-k-i}\Lambda_{i}.$$ 

We observe that $L_\Lambda+L_{xB}=n-k+1$. We easily see that $\Lambda^*(x)$ has degree up to $t_0$ if and only if $L_{\Lambda^*}=L_\Lambda$,
in particular, $\Lambda^*(x)=\Lambda(x)$ if $L_\Lambda<t_0$. This justifies Steps 1 and 2.  

Now assume $L_\Lambda=t_0$ and $\Delta_{i_1}=\Delta_{i_2}=\ldots=\Delta_{i_{t_0}}$. 
By letting $\Delta=\Delta_{i_1}$,  $\Lambda^*(x)$ has degree $t_0$, and at same time, contains $t_0$ valid roots, 
$\alpha^{-i_1}$, $\alpha^{-i_2}$, \ldots, $\alpha^{-i_{t_0}}$, i.e., $\Lambda^*(x)$ is a valid error locator polynomial. 
On the other hand, if an error locator polynomial ${\bar \Lambda}(x)$ is of degree $t_0$ and contains $t_0$ valid roots,
$\alpha^{-i_1}$, $\alpha^{-i_2}$, \ldots, $\alpha^{-i_{t_0}}$. 
Then, following \eqref{one-step}, we have
$$0=\Lambda(\alpha^{-i_l})-\Delta(S_{n-k})\cdot \alpha^{-i_l}B(\alpha^{-i_l}), \;\;\; l=1, 2, \ldots, t_0,$$
indicating $\Delta_{i_1}=\Delta_{i_2}=\ldots=\Delta_{i_{t_0}}=\Delta(S_{n-k})$. We thus justify Steps 3, 4, and 5.
\hfill $\Box\Box$

\noindent
{\bf Remarks:} Compared to the approach in \cite{Blahut}, where the syndrome $S_{n-k}$ is exhaustively searched throughout the field $\GF(q)$, 
each time $\Lambda^*(x)$ is produced and examined, the proposed algorithm reduces the computational complexity by a factor of $q$. 
Essentially, the proposed algorithm extends one iteration beyond the conventional Berlekamp-Massey algorithm 
while maintaining the original computational complexity. Further, note that an index (location) can only be classified to one group, 
thus any two sets of error locator roots are disjoint. 
As a result, there exist at most $\lfloor\frac{n}{t_0}\rfloor$ distinct codewords at distance $t_0$ from a received word. 
An extension of the above method is the $\lfloor\frac{d+1}{2}\rfloor$ decoding algorithm
which utilizes the Chien Search to determine the subsequent two unknown discrepancies  \cite{Egorov}.

{\bf Remark:} The foregoing one-step-ahead algorithm is essentially a degeneration of the list decoding algorithm to be presented in next section.

\subsection{Berlekamp Algorithm and its One-Step Extension for Decoding BCH Codes}

The underlying generator polynomial of a BCH code contains consecutive roots $\alpha$, $\alpha^2$, \ldots, $\alpha^{2t}$. 
Note for an underlying  binary BCH code, the designed minimum distance $d$ is always odd, 
which is actually a lower bound of the true minimum distance.

The Berlekamp algorithm is a simplified version of the Berlekamp-Massey algorithm for decoding binary BCH codes
by incorporating the special syndrome property 
 $$S_{2i+1}=S_{i}^2, \hspace{0.2in} i=0, 1, 2, \ldots$$
 which yields zero discrepancies at even iterations of the Berlekamp-Massey algorithm (cf. \cite{Berlekamp}). 
Below we re-formulate slightly the Berlekamp algorithm described in \cite{Berlekamp}, so as to facilitate the characterizations thereafter.

\vspace{0.1in}
{\underline {\bf Berlekamp Algorithm}}
{\small \begin{itemize}
\item Input:\hsp ${\bf S}=[S_0, \;\; S_1,\; \;S_2,\; \;\ldots,\;\; S_{d-2}]$
\item Initialization:\hsp $\Lambda^{(0)}(x)=1$, $B^{(-1)}(x)=x^{-1}$,  $L^{(0)}_\Lambda=0$, $L^{(-1)}_B=-1$
\item For $r=0$, 2, \ldots, $d-3$, \hsp do:
\begin{itemize}
\item Compute $\Delta^{(r+2)}=\sum_{i=0}^{L^{(r)}_\Lambda} \Lambda^{(r)}_i \cdot S_{r-i}$
\item Compute $\Lambda^{(r+2)}(x)=\Lambda^{(r)}(x) -\Delta^{(r+2)} \cdot x^2 B^{(r-1)}(x)$
\item If $\Delta^{(r+2)} \ne 0$ and $2L^{(r)}_\Lambda\leq r$, \hsp then
\begin{itemize}
\item Set $B^{(r+1)}(x)\gets (\Delta^{(r+2)})^{-1} \cdot \Lambda^{(r)}(x)$
\item Set $L^{(r+2)}_\Lambda\gets L^{(r-1)}_B+2$, \ \ $L^{(r+1)}_B\gets L^{(r)}_\Lambda$
\end{itemize}
\item Else 
\begin{itemize}
\item Set $B^{(r+1)}(x) \gets  x^2 B^{(r-1)}(x)$
\item Set $L^{(r+1)}_B\gets L^{(r-1)}_B+2$, \ \ $L^{(r+2)}_\Lambda\gets L^{(r)}_\Lambda$
\end{itemize}
\item[] endif
\end{itemize}
\item[] endfor
\item Output:\hsp $\Lambda(x)$, \ $B(x)$, \ $L_\Lambda$, \ $L_B$
\end{itemize}
}

The following lemma characterizes the error locator polynomial $\Lambda(x)$ 
and the correction polynomial $B(x)$ produced by the Berlekamp algorithm.
\begin{lemma}
Let $\Lambda(x)$ and $B(x)$ be the error locator and correction polynomials, respectively, computed by the Berlekamp algorithm.\\
$(i)$. $L_\Lambda$ and $L_B$, the length of AFSR described by $\Lambda(x)$ and $B(x)$ respectively,   satisfy 
\begin{equation}
L_\Lambda + L_B=d-2.
\end{equation}
$(ii)$. The degrees of $\Lambda(x)$ and $B(x)$ satisfy   
\begin{equation}
\deg(\Lambda(x)) \leq L_\Lambda, \hspace{0.3in} \deg(B(x)) \leq L_B.
\end{equation}
When $\Lambda(x)$ is the true error locator polynomial as defined in \eqref{def-true-Lambda}, $\deg(\Lambda(x))=L_\Lambda$. \\
$(iii)$. The polynomials  $\Lambda(x)$ and $B(x)$ are coprime, i.e., the two do not share a common factor. 
\end{lemma}

Similarly, we have the following one-step-ahead algorithm that corrects (in the list decoding sense) 
up to $\frac{d+1}{2}$ errors 
at essentially same complexity as the original Berlekamp algorithm. The proof is straightforward.

{\underline {\bf One-Step-Ahead Berlekamp Algorithm}}
\begin{enumerate}
\item If $L_\Lambda > \frac{d+1}{2} $, then declare a decoding  failure. 
\item If $L_\Lambda < \frac{d+1}{2} $, then determine all distinct roots in $\{\alpha^{-i}\}_{i=0}^{n-1}$.
If the number of (distinct) roots is equal to $L_\Lambda$, then return the corresponding unique codeword, 
otherwise declare a decoding failure (which is identical to the normal Berlekamp algorithm)
\item Evaluate $\Delta_i=\frac{\Lambda(\alpha^{-i})}{\alpha^{-2i}B(\alpha^{-i})}$, $i=0, 1, 2, \ldots, n-1$.
\item Group the index sets $\{i_1, i_2, \ldots, i_{L_\Lambda} \}$ such that $\Delta_i$'s are identical 
(each set corresponds to the roots of a valid error locator polynomial). 
\item Flip bits on all indices (locations) of each set obtained in Step 4, each resulting in a candidate codeword.
\end{enumerate}

\noindent
{\bf Remark:} The proposed algorithm is superior to the ``trick" presented in \cite{Duursma} 
in which $\frac{d+1}{2}$ error correction capability is achieved 
only for even-weight subcode by exploiting the parity syndrome of a received word.


\section{\sc List Decoding Algorithm for Reed-Solomon Codes}

In this section we present a list decoding algorithm for Reed-Solomon codes that corrects up to $\lceil n-1-\sqrt{n(n-d)}\rceil$ errors, 
which is identical to that of the Guruswami-Sudan algorithm in \cite{Guru-Sudan}. 
We shall extend the notation $t_0=\frac{d}{2}$ to allow $d$ to take any integer value, instead of even value as initially defined in \eqref{def-t0}. 
We use the terminology ``valid" root to indicate 
a root is in the pre-defined space which, in this context, means $\{1$, $\alpha^{-1}$, \ldots, $\alpha^{-(n-1)} \}$. 
We also define a companion polynomial $\bmQ(x, y)$ of a bivariate polynomial $\mQ(x, y)$ to be $\bmQ(x, y)\eqdef \mQ(x, 1/y)y^{P_y}$ 
(herein $P_y$ denotes the power of $y$ in $\mQ(x, y)$), and $[y^l]\mQ(x, y)$ the polynomial of $x$ associated with the term $y^l$. 
Specifically, 
\begin{eqnarray}
\bmQ(x, y) & \eqdef & f_{P_y}(x)+f_{P_y-1}(x) y + f_{P_y-2}(x) y^2 +\ldots + f_1(x) y^{P_y-1} + f_0(x) y^{P_y} \\
 \left[ y^l \right] \mQ(x, y) & \eqdef & f_l(x)
\end{eqnarray}
if $\mQ(x, y)$ is in form of $\mQ(x, y)=f_0(x)+ f_1(x) y + f_2(x) y^2+ \ldots+ f_{P_y}(x) y^{P_y}$.

\subsection{Problem Formulation}

\begin{lemma}   \label{LEM-Lambda-form}
Let $\Lambda^*(x)$ be the true error locator polynomial as defined in \eqref{def-true-Lambda}. Let $\Lambda(x)$ and $B(x)$ be the error locator and correction polynomials, respectively, 
obtained from the Berlekamp-Massey algorithm. Then, 
$\Lambda^*(x)$ exhibits the form of
\begin{equation}
\Lambda^*(x)= \Lambda(x) \cdot \lambda^*(x) + xB(x) \cdot b^*(x), \label{form-Lambda}
\end{equation}
where the polynomials $\lambda^*(x)$ and $b^*(x)$ exhibit the following properties\\
$(i)$. $\lambda^*_0=1$; \\
$(ii)$. if $b^*(x) = 0$, then $\lambda^*(x)= 1$; \\
$(iii)$. $\lambda^*(x)$ and $b^*(x)$ are coprime; \\
$(iv)$. $\deg(\lambda^*(x)) = \deg(\Lambda^*(x))-L_\Lambda$ \ \& \ $\deg(b^*(x))\leq \deg(\Lambda^*(x))-L_{xB}$, or \\
 $\deg(\lambda^*(x)) \leq  \deg(\Lambda^*(x))-L_\Lambda$ \ \& \ $\deg(b^*(x)) = \deg(\Lambda^*(x))-L_{xB}$; \\
$(v)$. if $\deg(\Lambda^*(x))<n-k$, then $\lambda^*(x)$ and $b^*(x)$ are unique.
\end{lemma} 
{\em Proof: } When the number of errors $e\leq\lfloor \frac{n-k}{2} \rfloor$, the above conclusions trivially hold, 
following the nature of the Berlekamp-Massey algorithm.
In the following we consider for the case $e>\lfloor \frac{n-k}{2} \rfloor$.
Suppose a genie tells additional syndromes $S_{n-k}, S_{n-k+1} \ldots, S_{2e-1}$,  
(alternative interpretation is to assume the all-zero codeword is transmitted and thus the additional syndromes are available), 
the true error locator polynomial $\Lambda^*(x)$ can be obtained by further applying the Berlekamp-Massey algorithm 
in conjunction with syndromes $S_{n-k}, S_{n-k+1}, \ldots, S_{2e-1}$.
Thus, \eqref{form-Lambda} holds by the nature of the Berlekamp-Massey algorithm. 

$(i)$. Note that $\Lambda^*_0(x)=1$ by definition \eqref{def-true-Lambda} and $\Lambda_0(x) =1$ by the nature of the Berlekamp-Massey iteration,
$\Lambda^{(r+1)}(x)=\Lambda^{(r)}(x)-\Delta^{(r+1)} xB^{(r)}(x)$, and by the initial condition $\Lambda^{(0)}(x)=1$.
On the other hand, the constant term of $xB(x) \cdot b^*(x)$ is always zero. Therefore, $\lambda^*_0(x)=1$.

$(ii)$. If $b^*(x)=0$, then the corresponding (additional) discrepancies $\Delta^{(n-k)}$,  $\Delta^{(n-k+1)}$, \ldots, $\Delta^{(2e)}$, are all zeros. 
We thus obtain 
$$\Lambda^*(x)=\Lambda^{(2e)}(x)=\Lambda^{(2e-1)}(x)=\ldots=\Lambda^{(n-k)}(x)=\Lambda(x),$$  
which justifies the property $(ii)$.

$(iii)$. Let $B^*(x)$ be the correction polynomial associated with $\Lambda^*(x)$.
It can be easily shown that (cf. \cite{Blahut})
$$
\left[\begin{array}{l} \Lambda^*(x) \\ B^*(x) \end{array} \right] =
\prod_{r=1}^{2e} \left[\begin{array}{ll}  1 & -\Delta^{(r)}x \\    (\Delta^{(r)})^{{-1}}\delta^{(r)} &  (1-\delta^{(r)})x \end{array}\right]
\left[ \begin{array}{l} 1 \\ 1 \end{array} \right]
$$
where $\delta^{(r)}$ denotes a binary value associated with selection of $B^{(r)}(x)$ and is zero when $\Delta^{(r)}=0$, and likewise,
$$
\left[\begin{array}{l} \Lambda(x) \\ B(x) \end{array} \right] =
\prod_{r=1}^{n-k} \left[\begin{array}{ll}  1 & -\Delta^{(r)}x \\    (\Delta^{(r)})^{{-1}}\delta^{(r)} &  (1-\delta^{(r)})x \end{array}\right]
\left[ \begin{array}{l} 1 \\ 1 \end{array} \right].
$$
The above equalities, in conjunction with \eqref{form-Lambda}, indicate that
$$
\left[\begin{array}{ll} \lambda^*(x) & x b^*(x) \end{array} \right] =
\left[ \begin{array}{ll} 1 & 0 \end{array} \right].
\prod_{r=n-k+1}^{2e} \left[\begin{array}{ll}  1 & -\Delta^{(r)}x \\    (\Delta^{(r)})^{{-1}}\delta^{(r)} &  (1-\delta^{(r)})x \end{array}\right].
$$
We proceed to show by induction that $\lambda^*(x)$ and $xb^*(x)$ are coprime.  
When $r=n-k+1$, we have $\lambda^{(1)}(x)=1$ and $xb^{(1)}(x)= -\Delta^{(n-k+1)}x$. Clearly $\lambda^{(1)}(x)$ and $xb^{(1)}(x)$ are coprime.
Assuming that $\lambda^{(i)}(x)$ and $xb^{(i)}(x)$ (i.e., the case $r=n-k+i$) are coprime, we then have 
\begin{eqnarray*}
\left[\lambda^{(i+1)}(x) \;\;\; xb^{(i+1)}(x) \right] \aline{=}
\left[ \lambda^{(i)}(x) \;\;\;  xb^{(i)}(x) \right] \cdot 
\left[\begin{array}{ll}  1 & -\Delta^{(r+1)}x \\    (\Delta^{(r+1)})^{{-1}}\delta^{(r+1)} &  (1-\delta^{(r+1)})x \end{array}\right] \\
\aline{=} \left[\lambda^{(i)}(x)+(\Delta^{(r+1)})^{{-1}}\delta^{(r+1)} xb^{(i)}(x)  \;\;
-\Delta^{(r+1)}x \lambda^{(i)}(x) + (1-\delta^{(r+1)}) x^2b^{(i)}(x) \right].
\end{eqnarray*}
When $\delta^{(r+1)}=0$, we obtain 
$$\left[\lambda^{(i+1)}(x) \;\;\; xb^{(i+1)}(x) \right] = \left[\lambda^{(i)}(x) \;\;\; -\Delta^{(r+1)}x \lambda^{(i)}(x) + x^2b^{(i)}(x) \right]
$$
which clearly indicates that $\lambda^{(i+1)}(x)$ and $xb^{(i+1)}(x)$ are coprime. When $\delta^{(r+1)}=1$, we obtain
$$\left[\lambda^{(i+1)}(x) \;\;\; xb^{(i+1)}(x) \right] = \left[\lambda^{(i)}(x)+(\Delta^{(r+1)})^{{-1}}xb^{(i)}(x) \;\;\; 
-\Delta^{(r+1)}x \lambda^{(i)}(x) \right]
$$
which again indicates that  $\lambda^{(i+1)}(x)$ and $xb^{(i+1)}(x)$ are coprime. 
Therefore, $\lambda^*(x)$ is coprime to $xb^*(x)$ and subsequently to $b^*(x)$.

$(iv)$. The results clearly hold when $\Lambda^*(x)=\Lambda(x)$. We next show for the case $\Lambda^*(x)\ne \Lambda(x)$, which indicates that
the number of errors $e> \frac{n-k}{2}$. Given the hypothetical $2e-(n-k)$ additional syndromes, $S_{n-k}$, $S_{n-k+1}$, \ldots, $S_{2e-1}$,
$\Lambda^*(x)$ is the minimum-length LFSR to generate the syndrome sequence $S_0$, $S_1$, $S_2$, \ldots, $S_{2e-1}$.
By Lemma~\ref{LEM-BMA-character}, $L_{\Lambda^*}=e=\deg(\Lambda^*(x))$. On the other hand, we observe that $\lambda^*(x)$ and $b^*(x)$ are obtained by 
further applying the Berlekamp-Massey iterations on top of  $\Lambda(x)=\Lambda^{(n-k)}(x)$ and $B(x)=B^{(n-k)}(x)$, 
whose LFSR lengths are $L_\Lambda$ and $L_B$, respectively.  
Therefore, we have 
\begin{eqnarray*}
\deg(\lambda^*(x)) \aline{ \leq}  L_{\Lambda^*} -L_\Lambda=\deg(\Lambda^*(x)) -L_\Lambda \\
\deg(xb^*(x)) \aline{\leq}  L_{\Lambda^*} -L_{B} = \deg(\Lambda^*(x))-L_{B}.
\end{eqnarray*}
Further note that \eqref{form-Lambda} indicates that 
$$\deg(\Lambda(x))+\deg(\lambda^*(x))=\deg(\Lambda^*(x)) \hspace{0.2in} \text{or} \hspace{0.2in}
\deg(xB(x))+\deg(b^*(x))=\deg(\Lambda^*(x)).$$
Without loss of generality, we assume $\deg(\Lambda(x))+\deg(\lambda^*(x))=\deg(\Lambda^*(x))$, which immediately yields
$$\deg(\lambda^*(x))= \deg(\Lambda^*(x))-\deg(\Lambda(x)) \geq \deg(\Lambda^*(x)) -L_\Lambda,$$
where the inequality follows Lemma~\ref{LEM-BMA-character}. We thus obtain
$$\deg(\lambda^*(x))=\deg(\Lambda^*(x)) -L_\Lambda  \text{ \ and \ } \deg(\Lambda(x))=L_\Lambda.$$

$(v)$. We prove Part $(v)$ by contradiction. Assume that there is another pair $\lambda(x)$ and $b(x)$ satisfying \eqref{form-Lambda}.
Then, we have
$$\Lambda^*(x) = \Lambda(x) \cdot \lambda^*(x) + xB(x) \cdot b^*(x) = \Lambda(x) \cdot \lambda(x) + xB(x) \cdot b(x),$$
which immediately indicates
$$\Lambda(x) \cdot (\lambda^*(x)-\lambda(x)) = xB(x) \cdot (b(x)- b^*(x)).$$
Since $\Lambda(x)$ is coprime to $xB(x)$ (by Lemma~\ref{LEM-BMA-character}.(iii)), 
$\Lambda(x)$ divides $b(x)- b^*(x)$. Likewise, $xB(x)$ divides $\lambda^*(x)-\lambda(x)$.
\eqref{form-Lambda} indicates that
$$ \deg(\Lambda^*(x))=\max\{ \deg(\lambda^*(x))+\deg(\Lambda(x)),\;\;  \deg(b^*(x))+\deg(xB(x)) \}$$
Without loss of generality, we assume $\deg(\Lambda^*(x))=\deg(\lambda^*(x))+\deg(\Lambda(x))$. Then,
$\deg (\Lambda(x)) = L_\Lambda $, following Part $(iv)$. Subsequently,
$$\deg (\Lambda(x)) =n-k-L_{B} > e-L_{B} \geq \deg(b^*(x)).$$
Likewise, $\deg (\Lambda(x)) > \deg(b(x))$. Therefore, $\Lambda(x)$ divides $b(x)- b^*(x)$ if and only if $b(x)- b^*(x)=0$, i.e., $b(x)=b^*(x)$. 
Part $(v)$ is thus justified.              \hfill $\Box\Box$

{\bf Example 1.} Consider a transmitted codeword pertaining to the (15, 5) Reed-Solomon code 
$$\bc=
[0,\;  \alpha^{12},\;  \alpha^{10},\;  \alpha^{11},\;  \alpha^{7},\;  \alpha^{5},\;  \alpha^{11},\;  \alpha^{11},\;  \alpha^{6},\;  \alpha^{8},\;  \alpha^{14},\; 
 \alpha^{11},\;  \alpha^{6},\;  \alpha^{2},\;  \alpha ] $$
where $\alpha$ denotes a primitive element of $\GF(16)$, and the corresponding received word which has 7 errors
$${\bf r}=[0,\;  \alpha^{6},\;  \alpha^{10},\;  \alpha^{5},\;  \alpha^{7},\;  \alpha^{5},\;  \alpha^{5},\;  \alpha^{11},\;  \alpha ,\;  \alpha^{2},\;  \alpha^{14},\; 
 \alpha^{12},\;  \alpha^{6},\;  \alpha^{7},\;  \alpha ].$$
The true error locator polynomial is
\begin{eqnarray*}
\Lambda^*(x) \aline{=} (1-\alpha x)(1-\alpha^{3}x)(1-\alpha^{6}x)(1-\alpha^{8}x)(1-\alpha^{9}x)(1-\alpha^{11}x)(1-\alpha^{13}x)  \\
\aline{=} 1+\alpha^{9} x+\alpha^{11} x^{3}+\alpha^{5} x^{4}+\alpha^{12} x^{6}+\alpha^{6} x^{7} .
\end{eqnarray*}
Applying the Berlekamp-Massey algorithm, we obtain the following error locator and correction polynomials
\begin{eqnarray*}
\Lambda(x) \aline{=} 1 + \alpha^7 x + \alpha^{13} x^2 + \alpha x^3 + \alpha^{13} x^4 + \alpha^4 x^5\\
B(x) \aline{=}  \alpha^6 x + \alpha^5 x^2 + \alpha^{10} x^3 + \alpha^3 x^4 + \alpha^{11} x^5.
\end{eqnarray*}
It can be verified that $\Lambda^*(x)$ satisfies the following decomposition 
$$\Lambda^*(x)=\Lambda(x)(1+ \alpha  x + \alpha^{2} x^{2} ) + xB(x)\cdot \alpha^{8}. $$
\hfill $\Box\Box$

\begin{lemma} \label{LEM-extreme-case}
Let $\Lambda(x)$ and $B(x)$ be the error locator and correction polynomials, respectively, 
obtained from the Berlekamp-Massey algorithm. Let the distance threshold $t\geq t_0$, where $t_0$ is defined in \eqref{def-t0}. \\
$(i)$. If the degree of $\Lambda(x)$, $L_\Lambda>t$, then there is no codeword within distance $t$ from the received word. \\
$(ii)$. If the degree of $xB(x)$, $L_{xB}>t$, then only $\Lambda^*(x)=\Lambda(x)$ may result in a codeword within distance $t$ from the received word.
\end{lemma}
{\em Proof: } We show $(i)$ by contradiction. Assume there is a codeword within distance $t$ from the received word. 
Then, the corresponding error locator locator polynomial $\Lambda^*(x)$, which has degree up to $t$, 
generates the syndrome sequence $S_0, S_1, \ldots, S_{n-k-1}$. This contradicts the fact that $\Lambda(x)$ 
represents a minimum-length shift register which generates $S_0, S_1, \ldots, S_{n-k-1}$.\\
$(ii)$. If $\Lambda(x)$ contains exactly $L_\Lambda$ distinct roots within $\{\alpha^{-i}\}_{i=0}^{n-1}$, 
then $\Lambda^*(x)=\Lambda(x)$ leads to a codeword with distance $L_\Lambda<t$ from the received word.
Assume there is a codeword at distance $e$ from the received word, where $e> L_{\Lambda}$.
Then, $e\geq t_0$. This is because, if $e<t_0$, then $\Lambda^*(x)$ is the unique minimum-length LFSR to generate 
the syndrome sequence, $S_0, S_1, \ldots, S_{n-k-1}$,
which obviously conflicts the facts $\Lambda^*(x) \ne \Lambda(x)$. Now assume that the additional syndromes 
$S_{n-k}$, $S_{n-k+1}$, \ldots, $S_{2e-1}$ are available and used for further applying the Berlekamp-Massey iterations. 
Let $\Delta^{(r)}$ ($r>n-k$)
be the first nonzero discrepancy, then the error locator polynomial is updated as 
$$\Lambda^{(r)}(x)=\Lambda^{(n-k)}(x)-\Delta^{(r)} x^{r-(n-k)} B^{(n-k)}(x),$$
which immediately indicates 
$$L_\Lambda^*\geq L^{(r)}_{\Lambda}=L_B+r-(n-k)\geq L_{xB}.$$
Since $\Lambda^*(x)$ is a true error locator polynomial, $e=L_{\Lambda^*}=\deg(\Lambda^*(x))$. 
Thus, $\Lambda^*(x)$ has degree at least $L_{xB}$, i.e., $e>L_{xB}$ and the proof is completed.
\hfill $\Box\Box$

{\bf Example 2.} Consider a transmitted codeword pertaining to the (15, 5) Reed-Solomon code
$$\bc=[\alpha^{2}, \; \alpha^{3},\; \alpha^{5} ,\;  \alpha^{9} ,\;  \alpha^{12} ,\;  \alpha^{5} ,\;  \alpha^{4} ,\;   \alpha  ,\;  \alpha^{9} ,\;  \alpha^{14} ,\;  \alpha^{9} ,\;  \alpha^{8} ,\;  
\alpha^{9} ,\;  \alpha^{11},\; \alpha^{10}]$$
and the corresponding received word
$${\bf r}=[\alpha^{2},\; \alpha^{8} ,\; \alpha^{5},\; \alpha^{9},\; 1,\; \alpha^{5},\; \alpha^{4},\; \alpha^{12} ,\; \alpha^{2},\; \alpha^{14}
,\; \alpha^{9},\; \alpha^{8} ,\; \alpha^{13},\; 1 ,\; \alpha^{11}]$$
which has 7 errors.
The true error locator polynomial is
\begin{eqnarray*} 
\Lambda^*(x)\aline{=}(1-\alpha x)(1-\alpha^{4}x)(1-\alpha^{7}x)(1-\alpha^{8}x)(1-\alpha^{12}x)(1-\alpha^{13}x)(1-\alpha^{14}x) \\
\aline{=} 1 +\alpha^{5} x +\alpha^{4} x^{3}+\alpha^{14} x^{4}+\alpha^{8} x^{5}+\alpha^{12} x^{6}+\alpha^{14} x^{7}.
\end{eqnarray*}
The Berlekamp-Massey algorithm returns the error locator and correction polynomials below
\begin{eqnarray*}
\Lambda(x) \aline{=} 1+ \alpha^9 x  + \alpha^5 x^2 + \alpha^6 x^3 + \alpha^9 x^5 + \alpha  x^6 + \alpha^4 x^7 \\
B(x) \aline{=} \alpha^{11} + \alpha^5 x  + \alpha  x^2 + \alpha^2 x^3 .
\end{eqnarray*}
Note that $\Lambda(x)$ has degree 7, Lemma~\ref{LEM-extreme-case} asserts no codewords 
within 6 symbol difference from the received word.  \hfill $\Box\Box$


\subsection{Rational Interpolation}

Dividing both sides of \eqref{form-Lambda} by  $xB(x)$, we obtain
$$\frac{\Lambda^*(x)}{xB(x)}= \frac{\Lambda(x)}{xB(x)} \cdot \lambda^*(x) + b^*(x).$$
Define
\begin{equation}
y_i=-\frac{\Lambda(\alpha^{-i})}{\alpha^{-i}B(\alpha^{-i})}, \hspace{0.3in}  i=0, 1, 2, \ldots, n-1 \label{def-y-value}
\end{equation}
where $y_i$ is set to $\infty$ when $B(\alpha^{-i})=0$, whose implication will be explored shortly.
Let $\alpha^{-i_1}$, $\alpha^{-i_2}$, \ldots, $\alpha^{-i_e}$, be all the valid roots of the true error locator polynomial $\Lambda^*(x)$. 
Then, $y\cdot \lambda^*(x)-b^*(x)$ passes precisely through $e$ points, $(\alpha^{-i_1}, \; y_{i_1})$, $(\alpha^{-i_2}, \; y_{i_2})$, \ldots, 
$(\alpha^{-i_e}, \; y_{i_e})$.

Given the set of $n$ distinct points $\{ (\alpha^{-i}, y_{i}) \}_{i=0}^{n-1}$,
we are interested in finding rational functions $y(x)$ which pass $t$ ($t\geq t_0$) points, in the sense that $y(\alpha^{-i})=y_i$. 
If $y_i=\infty$, then $y(x)$ must contain the pole $\alpha^{-i}$. This is because, when $B(\alpha^{-i})=0$, $\lambda(\alpha^{-i})$ must be zero, 
due to the fact that  $B(x)$ and $\Lambda(x)$ are coprime and thus cannot share the root.  
This is essentially a rational curve-fitting problem. 
In \cite{Guru-Sudan}, a powerful approach which makes use of multiple interpolation was presented to solve the polynomial curve-fitting problem. 
In essence, it constructs a global bivariate polynomial (curve) $\mQ(x, y)$ that passes through all points with certain multiplicity.
 By its algebraic nature all desired polynomials of the form $y-p(x)$ are its factors \cite{Guru-Sudan}.  
In the following we generalize the approach to the rational domain. 

The most efficient known interpolation technique was presented in \cite{Ma}, whose prototype 
was proposed in \cite{Koetter}. This approach exhibits quadratic complexity, as opposed to the straightforward Gaussian elimination method 
which exhibits cubic complexity.
We show that the same approach can also be applied to the rational interpolation with appropriate modifications. 
Firstly, the weight of $y$ is determined differently. 
Note that we are essentially interested in the form of $y\cdot \lambda(x)-b(x)$. We naturally assign the weight of $y$ to be
\begin{equation}
w \eqdef L_\Lambda-L_{xB}.  \label{def-w}
\end{equation}
We denote by $\deg_{1,w}(\mQ(x, y))$ the $(1, w)$-weighted degree of a bivariate polynomial $\mQ(x, y)$ 
(refer to \cite{Guru-Sudan, Koetter-Vardy} for a detailed description of ``weighted degree").
It is worth noting that the weight $w$ may take negative values, beyond the traditional notion.
Secondly, passing through the point $(\alpha^{-i}, \infty)$ with multiplicity $m$ has the special meaning, 
i.e., the companion polynomial $\bmQ(x, y)=\mQ(x, 1/y)y^{P_y}$ passes $(\alpha^{-i}, 0)$ at $m$ times. 
Finally, it is worth clarifying the (unconventional) relation between the power of $y$ in $\mQ(x, y)$, denoted by $P_y$, 
and the $(1, w)$-weighted degree of $\mQ(x, y)$, denoted by $\deg_{1, w}(\mQ(x, y))$. 
$P_y$ is no longer implicitly $\lfloor\frac{\deg_{1, w}(\mQ(x, y))}{w}\rfloor$ as in the case of polynomial interpolation, 
but a more sophisticated function  of $\deg_{1, w}(\mQ(x, y))$, as will be characterized in \eqref{Q-deg} in an optimal setup.

{\bf Example 3.} $(i)$. For the case presented in Example 1, the weight is set to $w=-1$. 
The $n=15$ interpolation points are
$$\begin{array}{lllll}
 (1,\;  \alpha^{6}) & (\alpha^{-1},\;  \alpha^{8}) &  (\alpha^{-2},\;  \alpha^{12}) &  (\alpha^{-3},\;  \alpha^{13}) &  (\alpha^{-4},\;  \alpha^{11}) \\
 (\alpha^{-5},\;  \alpha^{9}) & (\alpha^{-6},\;  \infty) & (\alpha^{-7},\;  \alpha^{4}) & (\alpha^{-8},\;  \alpha^{12}) & (\alpha^{-9},\;  \alpha^{13})  \\
(\alpha^{-10},\;  \alpha^{13}) & (\alpha^{-11},\;  \infty) &  (\alpha^{-12},\;  \alpha^{6}) & (\alpha^{-13},\;  \alpha^{10}) & (\alpha^{-14},\;  1) 
 \end{array} $$

The following $(1, -1)$-weighted degree bivariate polynomial passes each of the above points  7 times.
\begin{eqnarray*}
\aline{} \mQ(x, y) = \\
\aline{} \begin{array}{ll}
 y^{0}\cdot & [ \alpha^{7} + \alpha^{9} x  + \alpha^{14} x^{2} + \alpha^{5} x^{3} + \alpha^{2} x^{4} + \alpha^{8} x^{5} 
	+ \alpha^{12} x^{6} + \alpha^{6} x^{7} + \alpha  x^{8} + \alpha^{2} x^{9} + \alpha^{9} x^{10} + \alpha  x^{12} \\
&  + \alpha^{5} x^{13} + \alpha^{12} x^{14} + \alpha^{7} x^{15} ] \; + \\

 y^1  \cdot & [ \alpha^{11} + \alpha^{11} x  + \alpha^{14} x^{2} + \alpha^{9} x^{3} + \alpha^{12} x^{4} + \alpha^{11} x^{5} 
	+ \alpha^{11} x^{6} + x^{7} + \alpha^{4} x^{8} + \alpha^{7} x^{9} + \alpha^{9} x^{10} + \alpha^{11} x^{11} \\
& + \alpha^{8} x^{12} + \alpha^{13} x^{13} + \alpha^{14} x^{14} + \alpha^{10} x^{15} + \alpha^{2} x^{16} + \alpha  x^{17} ] \\

&  + \; \ldots \; + \; \ldots \; + \\

y^{15} \cdot & [  \alpha^{13} + \alpha^{14} x  + \alpha^{7} x^{2} + \alpha^{2} x^{3} + x^{4} + \alpha^{3} x^{5}
	+ \alpha^{5} x^{6} + \alpha^{6} x^{7} + \alpha^{9} x^{8} + x^{9} + \alpha^{2} x^{10} + \alpha^{14} x^{11} \\
& 	+ \alpha^{11} x^{12} + x^{13} + \alpha^{6} x^{14} + \alpha^{2} x^{15} + \alpha^{3} x^{16} + \alpha^{2}
	x^{17} + \alpha^{11} x^{18} + \alpha^{12} x^{19} + \alpha^{2} x^{20} + \alpha^{5} x^{21} \\
&	+ \alpha^{7} x^{22} + \alpha^{3} x^{23} + \alpha^{14} x^{24} + \alpha^{10} x^{25} + \alpha^{6} x^{26} + \alpha^{7} x^{27}
	+ \alpha  x^{28} + \alpha  x^{29} + \alpha^{2} x^{30} ] \; +\\

 y^{16} \cdot & [ \alpha^{6} + x  + \alpha^{12} x^{2} + \alpha^{12} x^{3} + \alpha^{6} x^{5} + \alpha^{7} x^{6}
	+ \alpha^{8} x^{7} + \alpha  x^{9} + \alpha^{3} x^{10} + \alpha^{9} x^{11} + \alpha^{4} x^{12} \\
&	+ \alpha^{9} x^{13} + \alpha^{8} x^{14} + \alpha^{4} x^{15} + \alpha^{2} x^{16} + \alpha  x^{17} + \alpha^{10} x^
	{18} + \alpha^{6} x^{19} + \alpha^{4} x^{20} + \alpha^{8} x^{21} + \alpha^{14} x^{22} \\
& 	+ \alpha^{9} x^{23} + \alpha^{11} x^{24} + \alpha^{11} x^{25} + \alpha^{9} x^{26} + x^{27} + \alpha^{6} x^{28} + \alpha^{10}
	x^{29} + \alpha^{14} x^{30} + \alpha^{2} x^{31} ].
\end{array}
\end{eqnarray*}
Note the $(1, -1)$--weighted degree of the above $\mQ(x, y)$ is 16, which is beyond the conventional notion of degree. 

$(ii)$. For the case given in Example 2, the weight is set to $w=3$. 
The $n=15$ interpolation points are
$$\begin{array}{lllll}
(1,  \; \infty )& (\alpha^{-1} , \;  \alpha^5) &  (\alpha^{-2} ,   \alpha^{2}) & (\alpha^{-3} ,   \alpha^{9})  & (\alpha^{-4} , \; \alpha^5)\\
 (\alpha^{-5} ,   \alpha^{4})  & (\alpha^{-6} ,   \alpha^{11})  & (\alpha^{-7} ,   \alpha)  & (\alpha^{-8} ,   \alpha^{4}) & (\alpha^{-9} ,   \alpha^{13}) \\
 (\alpha^{-10} ,   \alpha^{13}) & (\alpha^{-11} ,   \alpha) & (\alpha^{-12} ,   \alpha^{11}) & (\alpha^{-13} ,   \alpha^{3})  & (\alpha^{-14} , \;  \alpha^5)
 \end{array}$$
The following $(1, 3)$-weighted degree bivariate polynomial passes each of the above points  7 times.
\begin{eqnarray*}
\aline{} \mQ(x, y) = \\
\aline{} \begin{array}{ll}

 y^{0}\cdot  & [\alpha^4 + \alpha^{14} x + \alpha^2 x^2 + \alpha^9 x^3 + \alpha^{11} x^4 + \alpha^{14} x^6 + \alpha^{12} x^7 + 
	\alpha^4 x^8 + \alpha^8 x^9 + \alpha^2 x^{10} + \alpha^5 x^{11} + \alpha^{11} x^{12} \\
&  + \alpha^{14} x^{13} + \alpha^{10} x^{14} + \alpha^6
	x^{15} + \alpha^3 x^{16} + \alpha^{14} x^{17} + \alpha^{13} x^{18} + \alpha^5 x^{19} + \alpha^{12} x^{21} + x^{22} + \alpha^3 x^{23} + \alpha^{11} x^{24} \\
& + \alpha^8 x^{25} + \alpha^8 x^{26} + \alpha^{10} x^{28} + \alpha^6 x^{31} + \alpha^6 x^{32} + \alpha^7 x^{33}
	+ \alpha^{13} x^{34} + \alpha^5 x^{35} + \alpha  x^{36} + \alpha^{12} x^{37} + \alpha^2 x^{38}\\
&  + x^{39} + \alpha^4 x^{40} + \alpha^2x^{41} + \alpha^{10} x^{42} + \alpha^6 x^{43} + \alpha^7 x^{44} + \alpha^3 x^{45} 
	+ \alpha^{11} x^{46} + \alpha^5 x^{47} + \alpha^3x^{48}  ] \; +\\

y^1 \cdot & [\alpha^8 + \alpha^7 x + \alpha^3 x^2 + \alpha^{14} x^3 + \alpha^{10} x^4 + \alpha^9 x^5 + \alpha^3 x^6 + \alpha^8
	x^7 + \alpha  x^{10} + \alpha^{14} x^{11} + \alpha^6 x^{12} + \alpha^4 x^{13}\\
&  + \alpha^6 x^{14} + \alpha^4 x^{15} + \alpha^6 x^{16} + \alpha^4 x^{17} + \alpha^9 x^{18} + \alpha^7 x^{19} + x^{20} + \alpha^{10} x^{21} + \alpha^{13} x^{22} + \alpha^9 x^{23} \\
& + \alpha^9 x^{24} + \alpha^5 x^{25} + \alpha^{12} x^{26} + \alpha^{10} x^{27} + \alpha^{12} x^{28} + \alpha^4 x^{29} + \alpha^{10} x^{30} +
	\alpha^{11} x^{31} + \alpha^7 x^{32} + \alpha^6 x^{34}\\
& + \alpha^{10} x^{35} + x^{36} + \alpha^3 x^{37} + \alpha^5 x^{38} + \alpha^2 x
	^{39} + \alpha^{10} x^{40} + \alpha^3 x^{41} + \alpha  x^{42} + \alpha^8 x^{43} + \alpha^6 x^{45}]\\

& +\; \ldots\; + \; \ldots \; + \\

y^{12} \cdot &  [ 1 + \alpha^4 x^2 + \alpha^4 x^3 + \alpha^6 x^4 + \alpha^4 x^6 + \alpha^{14} x^7 + \alpha^{12} x^8 + \alpha^{11} x^9 + \alpha^{11} x^{10} + \alpha^3 x^{11}  ] \; +\\

y^{13} \cdot & [ \alpha^{11} x + \alpha^{10} x^2 + \alpha^{13} x^3 + \alpha^5 x^4 + \alpha^{11} x^5 + \alpha^{10} x^6 + \alpha^{13} x^7 + \alpha^5 x^8 ]
\end{array}
\end{eqnarray*}
								\hfill $\Box\Box$

\begin{lemma}  \label{LEM-zero-constraint}
Let $\mQ(x, y)$ be a bivariate polynomial passing through all $n$ points
$\{ (1, y_0)$, $(\alpha^{-1}, y_1)$,  $(\alpha^{-2}, y_2)$, \ldots, $(\alpha^{-(n-1)}, y_{n-1}) \}$, 
where $y_i$ is defined in \eqref{def-y-value}, each with multiplicity $m$.
If
\begin{equation}
(t-L_\Lambda)P_y + \deg_{1, w}(\mQ(x, y))  < tm,  \label{zero-constraint}
\end{equation}
where $P_y$ denotes the power of $y$ in $\mQ(x, y)$ and $\deg_{1, w}(\mQ(x, y))$ (where $w$ is defined in \eqref{def-w}) 
denotes the $(1, w)$-weighted degree of $\mQ(x, y)$, 
then $\mQ(x, y)$ contains all factors of the form $y\lambda(x)-b(x)$ which pass through $t$ ($t\geq L_\Lambda$) points.
\end{lemma}
{\em Proof: } Let $y\lambda(x) - b(x)$ be a polynomial passing through $t$ points. Then, 
$$g(x)\eqdef \lambda^{P_y}(x) \mQ\left(x, \;\; \frac{b(x)}{\lambda(x)}\right)$$
is a polynomial and contains at least $tm$ roots, i.e., all roots of $\Lambda^*(x)$ each with multiplicity $m$. 
We proceed to show that the degree of $g(x)$ is at most $(t-L_\Lambda)P_y + \deg_{1, w}(\mQ(x, y))$.
This is true for the starting term without involving $y$, i.e., $\lambda^{P_y}(x)\mQ(x, 0)$, following the fact
$$ \deg \left( \lambda^{P_y}(x)\mQ(x, 0) \right) \leq \deg(\lambda(x)) P_y + \deg_{1, w}(\mQ(x, y)) \leq  (t-L_\Lambda)P_y + \deg_{1, w}(\mQ(x, y)),$$
where $\deg(\lambda(x)) \leq (t-L_\Lambda)$ is due to Lemma~\ref{LEM-Lambda-form}.$(iv)$.
It also holds true for the ending term associated with $y^{P_y}$, i.e.,
$b^{P_y}(x) \cdot [y^{P_y}]\mQ(x, y)$, as follows
\begin{eqnarray*}
\deg\left(  b^{P_y}(x) \cdot [y^{P_y}]\mQ(x, y) \right) \aline{\leq}  \deg(b(x))P_y+(\deg_{1, w}(\mQ(x, y))-wP_y) \\
\aline{=} \deg_{1, w}(\mQ(x, y))+P_y(\deg(b(x))+L_{xB}-L_\Lambda) \\
\aline{\leq} \deg_{1, w}(\mQ(x, y))+P_y(t-L_\Lambda),
\end{eqnarray*}
where $\deg(b(x))+L_{xB} \leq t$ is due to Lemma~\ref{LEM-Lambda-form}.$(iv)$.
Hence, it trivially holds true for the intermediate terms. Finally, the fact 
$$\deg(g(x)) \leq (t-L_\Lambda)P_y + \deg_{1, w}(\mQ(x, y))<tm$$ 
indicates the polynomial $g(x)$ has more roots than its degree. This is possible only with $g(x)\equiv 0$. Therefore, we conclude that $y\cdot\lambda(x)-b(x)$ divides $\mQ(x, y)$. \hfill $\Box\Box$

On the other hand, a sufficient condition for $\mQ(x, y)$ to pass through all $n$ points
$\{ (1, y_0)$, $(\alpha^{-1}, y_1)$,  $(\alpha^{-2}, y_2)$, \ldots, $(\alpha^{-(n-1)}, y_{n-1}) \}$, each with multiplicity $m$, 
is that the number of coefficients (degrees of freedom) of $\mQ(x, y)$ is greater than the number of linear constraints.
Note that the degrees of freedom, denoted by $N_\text{free}$, is easily seen to be
\begin{equation}
N_\text{free}=\sum_{i=0}^{P_y} (\deg_{1, w}(\mQ(x, y))+1-iw) = \frac{(2\deg_{1, w}(\mQ(x, y))+2-wP_y)(P_y+1)}{2} 
\end{equation}
whereas passing through a point with multiplicity $m$ results in $\frac{m(m+1)}{2}$ linearly independent constraints 
(readers are referred to \cite{Guru-Sudan} for detailed description). Thus, the number of linear constraints, denoted by $N_\text{cstr}$, is
\begin{equation}
N_ \text{cstr}= \frac{nm(m+1)}{2} .
\end{equation}

We proceed to maximize the degrees of freedom subject to 
fixed number of errors $e=t$ and the fixed multiplicity $m$ and the constraint \eqref{zero-constraint}, as follows
\begin{eqnarray}
N_\text{free}&=& \frac{(2\deg_{1, w}(\mQ(x, y))+2-wP_y)(P_y+1)}{2} \nonumber \\
&\leq& \frac{(2(tm-1-(t-L_\Lambda)P_y)+2-wP_y)(P_y+1)}{2} \nonumber \\
&=& (tm-P_y(t-t_0))(P_y+1) \nonumber \\
&=& -(t-t_0)P_y^2 +P_y(tm-t+t_0) +tm  \nonumber \\
&=& -(t-t_0) \left(P_y-\frac{tm-t+t_0}{2t-2t_0} \right)^2  + \frac{ ( tm+t-t_0 )^2 }{ 4(t-t_0) } \label{max-freedoms} 
\end{eqnarray}
where ``=" is achieved in ``$\leq$" if and only if
\begin{equation}
\deg_{1, w}(\mQ^*(x, y))  = tm-1 - (t-L_\Lambda)P_y \label{Q-deg}
\end{equation}
which optimally accommodates the zero constraint \eqref{zero-constraint}, and the first equality is due to the fact
$$L_\Lambda-\frac{w}{2}=\frac{L_\Lambda+L_{xB}}{2}=\frac{n-k+1}{2}=t_0.$$

Clearly, the maximum degrees of freedom is achieved by choosing $P_y^*$ to be the closest integer to $\frac{tm-t+t_0}{2t-2t_0}$, 
 i.e.,
\begin{equation}
P_y^*=\bigg\lfloor \frac{tm-t+t_0}{2t-2t_0} +0.5 \bigg\rfloor = \bigg\lfloor \frac{tm}{2t-2t_0}\bigg\rfloor . \label{y-deg}
\end{equation}
Therefore, the optimal choice of $m$ is the minimum integer that enforces $N_\text{free}>N_\text{cstr}$, i.e., 
\begin{equation}
 -(t-t_0) \left( \bigg\lfloor \frac{tm}{2t-2t_0}\bigg\rfloor  -\frac{tm-t+t_0}{2t-2t_0} \right)^2  + \frac{ ( tm+t-t_0 )^2 }{ 4(t-t_0) }   
> \frac{nm(m+1)}{2} . \label{org-free-cstr}
\end{equation}

We next present an explicit construction of a valid (but not necessary optimal) multiplicity.  
Note the maximum degrees of freedom is bounded by
\begin{eqnarray}
\max\{N_\text{free}\}  
\aline{=} -(t-t_0) \left( \bigg\lfloor \frac{tm}{2t-2t_0}\bigg\rfloor  -\frac{tm-t+t_0}{2t-2t_0} \right)^2  + \frac{ ( tm+t-t_0 )^2 }{ 4(t-t_0) } \nonumber \\
\aline{\geq} -\frac{t-t_0}{4} +  \frac{ ( tm+t-t_0 )^2 }{ 4(t-t_0) }. 
\end{eqnarray}
Hence, to solve for the linear equation system, it suffices to enforce
\begin{equation}
\frac{ ( tm+t-t_0 )^2 }{ 4(t-t_0) } -\frac{t-t_0}{4}  > \frac{nm(m+1)}{2},
\end{equation}
which is reduced to
\begin{equation}
m^2(t^2-2n(t-t_0)) -2m(t-t_0)(n-t) > 0.  \label{m-t-constraint}
\end{equation}
The above inequality holds true for sufficiently large $m$ if and only if 
$$t^2-2n(t-t_0)>0$$
which in turn indicates that 
\begin{equation}
t<n-\sqrt{n(n-d)}  \label{t-limit}
\end{equation}
which is equal to that of \cite{Guru-Sudan}. 
When $t<n-\sqrt{n(n-d)}$,  it suffices to choose the multiplicity $m$ to be
\begin{equation}
m^*=\bigg\lfloor \frac{2(t-t_0)(n-t)}{t^2-2n(t-t_0)}+1  \bigg\rfloor = \bigg\lfloor \frac{t(2t_0-t)}{t^2-2n(t-t_0)}  \bigg\rfloor.  \label{m-bound}
\end{equation} 

The following derives a lower bound on the optimal multiplicity $m_\text{opt}$ by relaxing $P_y$ to a real number. More specifically,
 with $P_y$ being a real number, the constraint $N_\text{free}>N_\text{cstr}$  simplifies to
\begin{eqnarray*} 
&& \frac{ ( tm+t-t_0 )^2 }{ 4(t-t_0) }   > \frac{nm(m+1)}{2}        \\
&\Leftrightarrow& m^2(t^2-2n(t-t_0)) -2m(t-t_0)(n-t) +(t-t_0)^2 > 0
\end{eqnarray*}
which  indicates 
\begin{eqnarray}
m&> &\frac{(t-t_0)(n-t)+\sqrt{(t-t_0)^2(n-t)^2 - (t^2-2n(t-t_0))(t-t_0)^2} } { t^2-2n(t-t_0) } \nonumber \\ 
&=&\frac{(t-t_0)\left(n-t+\sqrt{n(n-d)}\right) }  { t^2-2n(t-t_0) }  \nonumber \\
&=&\frac{t-t_0}{n-\sqrt{n(n-d)}-t}.  \label{m-lowerbound}
\end{eqnarray}
We thus conclude that the optimal value of $m$ is within the range 
\begin{equation}
\bigg\lfloor 1+\frac{t-t_0}{n-\sqrt{n(n-d)}-t} \bigg\rfloor \leq m_\text{opt} \leq \bigg\lfloor \frac{t(2t_0-t)}{t^2-2n(t-t_0)}  \bigg\rfloor. \label{m-range}
\end{equation}

It is worth noting that the LECC $t$, the multiplicity $m$ and the $y$-degree $P_y$  are all irrelevant to
the degrees of $\Lambda(x)$ or $B(x)$ individually. 
We ought to ensure there is no negative freedom terms with the above choice, 
which is easily verified by the fact that $\deg_{1, w}(\mQ(x, y))+1-iw$ is always positive for $i=0, 1, 2, \ldots, P_y$.

We summarize the above discussions into the following lemma.
\begin{lemma}
Let $t$ satisfy \eqref{t-limit},  the multiplicity $m$ be chosen as in 
\eqref{m-bound}  and the $y$-degree $P_y$ of bivariate polynomial $\mQ(x, y)$ as in \eqref{y-deg}.  Then, all valid polynomials $\lambda(x) \cdot y - b(x)$ that pass through
exactly $t$ points are factors of the minimum $(1, \; w)$-weighted (where $w$ is defined in \eqref{def-w}) degree 
polynomial $\mQ(x, y)$ 
that passes through $\{(\alpha^{-i}, \; y_i) \}_{i=0}^{n-1}$ (where $y_i$ is defined in \eqref{def-y-value}), each with multiplicity $m$.  
\end{lemma}

However, different values of $e$ require different values of $P_y$ and $m$. We next show that for a given LECC $t$, 
it suffices to choose a unified value $P_y$ and $m$ based upon the LECC $t$ to identify 
all valid polynomials $\lambda(x) \cdot y - b(x)$ corresponding to {\em up to} $t$ errors.
Since the degrees of freedom is independent of the actual number of errors,  it is  only left to show that
$$(e-L_\Lambda)P_y+\deg_{1, w}(\mQ(x, y))=(e+t-2t_0)P_y+(t-t_0)<em,$$
where $\deg_{1, w}(\mQ(x, y))$ satisfies \eqref{Q-deg} and $m$ satisfies \eqref{m-lowerbound}.
By assumption the above inequality holds when $e=t$, i.e.,
$$\frac{(2t-2t_0)P_y}{tm-t+t_0} <1.$$
Therefore, it suffices to show 
\begin{eqnarray*}
&& \frac{(2t-2t_0)P_y}{tm-t+t_0} \geq  \frac{(t+e-2t_0)P_y}{em-t+t_0} \\
 & \Leftrightarrow & (t-e)P_y \left( (t-2t_0)m+(t-t_0) \right)\leq 0 \\
& \Leftrightarrow & m\geq\frac{t-t_0}{2t_0-t} ,
\end{eqnarray*}
which holds true following \eqref{m-lowerbound} and the fact $2t_0=d>n-\sqrt{n(n-d)}$.

The following theorem wraps up the subsection.
\begin{theorem}
Let $\Lambda(x)$ and $B(x)$ be the error locator and correction polynomials, respectively, computed by the Berlekamp-Massey algorithm.
For any given $t$ satisfying \eqref{t-limit}, if we choose the multiplicity $m$ as in \eqref{m-bound} and the $y$-degree $P_y$ 
of bivariate polynomial $\mQ(x, y)$ as in \eqref{y-deg},
 then the minimum $(1, L_\Lambda-L_{xB})$-weighted degree polynomial  $\mQ(x, y)$ that passes through each of $n$  distinct points, 
$\{(\alpha^{-i}, \; y_i) \}_{i=0}^{n-1}$ (where $y_i$ is defined in \eqref{def-y-value}), with multiplicity $m$, 
contains all factors $\lambda(x) y -b(x)$
 which pass through at least $L_\Lambda$ but at most $t$ out of $n$ points.
\end{theorem}
 
\noindent
{\bf Remarks:} The above theorem can be viewed as a generalization of Lemma~5 in \cite{Guru-Sudan} from polynomial to rational curve-fitting. 
Although we do not address this explicitly, the choice of weight of $y$ in $\mQ(x, y)$ is optimal 
in the sense of maximizing the LECC $t$. It optimally trades off 
between \eqref{zero-constraint} and \eqref{m-t-constraint}.

{\bf Example 4.} For the (15, 5) Reed-Solomon code, we have 
$n-\sqrt{n(n-d)}=7.254.$
We consider correcting up to $t=7$ errors. It suffices to choose the multiplicity  
$m=\big\lfloor \frac{t(2t_0-t)}{t^2-2n(t-t_0)} \big\rfloor=7.$
Let the degree of $y$ be chosen as $P_y=\big\lfloor\frac{tm}{2t-2t_0} \big\rfloor=16$, following \eqref{y-deg},
and the degree of $\mQ(x, y)$, $\deg_{1, w}(\mQ(x, y))$, as in \eqref{Q-deg}.
As a result, the degrees of freedom is $(tm-P_y(t-t_0))(P_y+1)=425$
whereas the number of linear constraints is $nm(m+1)/2=420.$
The lower bound of $m$ is $\big\lfloor 1+\frac{t-t_0}{n-\sqrt{n(n-d)}-t} \big\rfloor=6.$
When $m=6$, it can be verified that the maximum degrees of freedom and the number of constraints are both 315.
Thus, $m=6$ is not satisfactory and the optimal choice of multiplicity is $m=7$.
For the received words given in Examples 1 \& 2, the minimum (weighted) degree bivariate polynomials $\mQ(x, y)$ 
constructed with $m=7$ and $P_y=16$ are presented in Example 3.  
In contrast, to achieve the same LECC $t=7$, 
the Guruswami-Sudan algorithm requires multiplicity $m=16$ with $P_y=31$ (recall that $P_y$ is a natural upper bound on the list size). \hfill $\Box\Box$


\subsection{Rational Factorization}

In this section we apply rational factorization to obtain $\frac{b(x)}{\lambda(x)}$, following the developments in \cite{Roth, Wu-Siegel}. 
However, our particular application is complicated by not knowing {\em a priori} 
the degrees of  $\lambda(x)$ and $b(x)$.

Define 
\begin{equation}
f(x) \eqdef \frac{b(x)}{\lambda(x)}
\end{equation}
which is well defined, as indicated by Lemma~\ref{LEM-Lambda-form} that $b(x)$ and $\lambda(x)$ are coprime. 
We, thus, only need to determine $f(x)$ to construct $\Lambda^*(x)$.
Note that $\lambda_0=1$, as indicated by Lemma~\ref{LEM-Lambda-form}.
Consequently, $f(x)$ can be expressed  by an infinite-length non-negative power series
$$f(x) = \frac{b(x)}{\lambda(x)} = s_0+s_1 x +s_2 x^2+\ldots s_i x^i +\ldots.$$

When $\mQ(x, y)$ contains factors of the form $y-f(x)$, 
we can obtain the infinite-length power series of the rational function $f(x)$ 
through making use of the following factorization procedure \cite{Roth, Wu-Siegel}.

\vspace{0.1in}
{\underline {\bf Rational Factorization Procedure}}
\begin{enumerate}
\item[0.] Initialization: \ $i=0$ and $\mQ^{(0)}(x, y)\gets \mQ(x, y)$.
\item Determine the roots of $\mQ^{(i)}(0, y)$. 
\item For each root $s$, 
\begin{itemize}
\item Compute the shifted polynomial: \  $\hmQ^{(i)}(x, y) \gets \mQ^{(i)}(x, s+y)$.
\item Transform and then remove $x$ factors: \  $\mQ^{(i+1)}(x, y)\gets \hmQ^{(i)}(x, xy)/x^a$, 
where $x^a$ denotes the largest power of $x$ contained as a factor in $\hmQ^{(i)}(x, xy)$.
\end{itemize} 
\item Set $i\gets i+1$ and repeat Steps 1 and 2 for the derivative $\mQ^{(i)}(x, y)$ with respect to each root $s$.
\end{enumerate}

We proceed to show that a pertinent finite-length, say $L_s$,  power series of $s(x)$ suffices to retrieve $b(x)$ and $\lambda(x)$. 
Note we have
\begin{equation}
\frac{b(x)}{\lambda(x)} \equiv s(x) \;\; \pmod{x^{L_s}}.
\end{equation}
It can be efficiently solved by the Berlekamp-Massey algorithm given that $L_s$ is large enough, as demonstrated below.

We first assume that the degrees of $b(x)$ and $\lambda(x)$ are known {\em a priori}.
When $\deg(b(x))<\deg(\lambda(x))$,  
$\lambda(x)$ is then uniquely determined (upon normalization) 
by the Berlekamp-Massey algorithm (note that $s(x) \pmod{x^{L_s}}$, $\lambda(x)$, and $b(x)$ correspond to 
syndrome, error locator, and error evaluator polynomials, respectively),
with $L_s = 2\deg(\lambda(x))$. Note that the condition $\deg(b(x))<\deg(\lambda(x))$ implies
$$\sum_{j=0}^{\deg(\lambda(x))} \lambda_j s_{i-j}=0, \hspace{0.3in} i=\deg(\lambda(x)),\; \deg(\lambda(x))+1, \;\ldots, \; 2\deg(\lambda(x))-1.$$
Therefore, the LFSR length described by $\lambda(x)$ is $L_\lambda=\deg(\lambda(x))$.

When $\deg(b(x))\ge \deg(\lambda(x))$, let 
$$\frac{b(x)}{\lambda(x)}=a(x)+\frac{b'(x)}{\lambda(x)}$$
where $a(x)$ is a polynomial of degree $\deg(b(x))-\deg(\lambda(x))$ and the degree of $b'(x)$ is less than that of $\lambda(x)$. 
Let $j_1=\deg(b(x))-\deg(\lambda(x))$ and $j_2=\deg(b(x))+\deg(\lambda(x))$. 
We observe that $\lambda(x)$ generates the sequence $(s_0-a_0)$, \ldots, $(s_{j_1}-a_{j_1})$, 
$s_{j_1+1}$, \ldots, $s_{j_2}$,
consequently, its subsequence $s_{j_1+1}$, $s_{j_1+2}$, \ldots, $s_{j_2}$.
Therefore,  $\lambda(x)$ can be uniquely determined (upon normalization) using the sequence 
$s_{j_1+1}$, $s_{j_1+2}$, \ldots, $s_{j_2}$, 
and subsequently $b(x)$ is computed by following the equality
$$b(x)=s(x)\lambda(x) \;\; \pmod{x^{\deg(b(x))+1}}.$$
Clearly, in this case it suffices to compute the power series expansion of $f(x)$ up to length $L_s=\deg(b(x))+\deg(\lambda(x))+1$.

Because we do not know {\em a priori} the degrees $\deg(b(x))$ and $\deg(\lambda(x))$, we have to take into account the worst cases. 
When $\deg(b(x))=t-L_{xB}$ and $\deg(\lambda(x))=0$, the Berlekamp-Massey algorithm has to start from the syndrome $s_{t-L_{xB}+1}$, 
on the other hand, $\deg(\lambda(x))$ may be as large as $t-L_\Lambda$, 
hence the Berlekamp-Massey algorithm requires $2(t-L_\Lambda)$ syndromes in the worst case. 
Therefore we may safely set the length of $s(x)$ to:
\begin{equation}
L_s=t-L_{xB}+1 + 2(t-L_\Lambda)=3t+1-2t_0-L_\Lambda.
\end{equation}
We thus obtain $\lambda(x)$ by applying the Berlekamp-Massey algorithm 
which starts from the syndrome $s_{t-L_{xB}+1}$ and iterates $2(t-L_\Lambda)$ times. 
Thereafter, we compute $b(x)$ via
\begin{equation}
b(x)=s(x)\lambda(x) \;\; \pmod{x^{t-L_{xB}+1}}.
\end{equation}

After we have determined the error locator polynomial $\Lambda'(x)=\Lambda(x)\lambda(x)+xB(x)b(x)$, 
we can subsequently identify all error locations 
and apply Forney's formula to compute the corresponding error magnitudes \cite{Blahut}.

{\bf Example 5.} $(i)$. For the bivariate polynomial $\mQ(x, y)$ constructed in Example 3.$(i)$,  applying the proposed 
rational factorization returns three candidate rational functions.\\
(1). $\lambda(x)= 1+ \alpha  x + \alpha^{2} x^{2} $ and $b(x)= \alpha+ \alpha^{2} x $.
The candidate error locator polynomial is constructed 
\begin{eqnarray*}
\Lambda'(x) \aline{=} \lambda(x)\cdot\Lambda(x)+b(x)\cdot xB(x) \\
\aline{=} 1 + \alpha^9 x + \alpha^5 x^2 + \alpha^{10} x^3 + \alpha^4 x^4 + \alpha^7 x^5 + x^6 + \alpha^4 x^7 \\
\aline{=} (1-x)(1-\alpha^{3}x)(1-\alpha^{6}x)(1-\alpha^{7}x)(1-\alpha^{10}x)(1-\alpha^{11}x)(1-\alpha^{12}x)
\end{eqnarray*}
which produces the following candidate codeword
$$\bc=[\alpha^{2},\;  \alpha^{6},\;  \alpha^{10},\;  \alpha^{8},\;  \alpha^{7},\;  \alpha^{5},\;  \alpha^{6},\;  \alpha^{12},\;  \alpha ,\;  
\alpha^{2},\;  \alpha^{3},\;  \alpha^{9},\;  \alpha^{9},\;  \alpha^{7},\;  \alpha ]. $$
(2). $\lambda(x) = 1+ \alpha^{12} x  $ and $b(x) = \alpha^{7}$.
The candidate error locator polynomial is constructed as
$$\Lambda'(x) =1 + \alpha^2 x^1 + \alpha^4 x^2 + \alpha^5 x^3 + \alpha^2 x^4 + \alpha^4 x^5 + \alpha^{10} x^6$$
which has less than 6 distinct roots in $\GF(16)$ and thus is spurious. \\
(3). $\lambda(x) = 1+ \alpha  x + \alpha^{2} x^{2}$ and $b(x) = \alpha^8$.
The corresponding candidate error locator polynomial is 
\begin{eqnarray*}
\Lambda'(x) \aline{=} 1 + \alpha^9 x^1 + \alpha^{11} x^3 + \alpha^5 x^4 + \alpha^{12} x^6 + \alpha^6 x^7 \\
\aline{=} (1-\alpha x)(1-\alpha^{3}x)(1-\alpha^{6}x)(1-\alpha^{8}x)(1-\alpha^{9}x)(1-\alpha^{11}x)(1-\alpha^{13}x)
\end{eqnarray*}
which successfully retrieves the transmitted codeword
$$\bc=[0,\;  \alpha^{12},\;  \alpha^{10},\;  \alpha^{11},\;  \alpha^{7},\;  \alpha^{5},\;  \alpha^{11},\;  \alpha^{11},\;  \alpha^{6},\; 
 \alpha^{8},\;  \alpha^{14},\;  \alpha^{11},\;  \alpha^{6},\;  \alpha^{2},\;  \alpha ].$$

$(ii)$. For the bivariate polynomial $\mQ(x, y)$ constructed in Example 3.$(ii)$,  applying the proposed 
rational factorization returns one candidate rational function: $\lambda(x)=1$ and 
$b(x)= \alpha^{12}+ \alpha^{10} x+ \alpha^{4} x^{2}+ \alpha^{7} x^{3}$. Consequently,
the candidate error locator polynomial is constructed
\begin{eqnarray*}
\Lambda'(x) \aline{=} \lambda(x)\cdot\Lambda(x)+b(x)\cdot xB(x) \\
\aline{=}  1 + \alpha^5 x  + \alpha^4 x^3 + \alpha^{14} x^4 + \alpha^8 x^5 + \alpha^{12} x^6 + \alpha^{14} x^7 \\
\aline{=}(1-\alpha x)(1-\alpha^{4}x)(1-\alpha^{7}x)(1-\alpha^{8}x)(1-\alpha^{12}x)(1-\alpha^{13}x)(1-\alpha^{14}x)
\end{eqnarray*}
which produces the original codeword.
It also verifies Lemma~\ref{LEM-extreme-case}.$(i)$, that there does not exist codewords within distance 6 from the received word. \hfill $\Box\Box$


\subsection{Algorithmic Description and Performance Assertion}

We summarize the complete list decoding algorithm and its characterization as follows.\\[0.1in]

{\large \bf \underline {List Decoding Algorithm for Reed-Solomon Codes}}
\begin{enumerate} 
\item[0.] Initialization: Input code length $n$, information dimension $k$, 
and LECC $t$ satisfying $t<n-\sqrt{n(n-d)}$. 
Initialize the multiplicity $m$ based on \eqref{m-bound}, then greedily optimize it subject to the constraint \eqref{org-free-cstr}, 
subsequently  choose the $y$-degree $P_y$ as in \eqref{y-deg}. 

\item Input the received word and compute syndromes.

\item Apply the Berlekamp-Massey algorithm to determine $\Lambda(x)$ and $B(x)$. If $L_\Lambda >t$ then declare a decoding  failure.

\item Perform error correction and  evaluate $y_i=-\frac{\Lambda(\alpha^{-i})} { \alpha^{-i}B(\alpha^{-i}) } $, $i=0, 1, 2, \ldots, n-1$.

\item If $L_{xB}>t$,  then return the corresponding unique codeword when $\Lambda(x)$ has $L_\Lambda$ valid roots, 
otherwise declare a decoding  failure.

\item Apply the rational interpolation procedure to compute a (1, $L_\Lambda-L_{xB}$)-weighted-degree polynomial $\mQ(x, y)$ 
that passes through $\left\{(\alpha^{-i}, \; y_i)\right\}_{i=0}^{n-1}$, each with multiplicity $m$.

\item Apply the rational factorization process to obtain finite-length power series $s(x)\;\; \pmod{x^{L_s}}$ 
of the rational functions $\frac{b(x)}{\lambda(x)}$.

\item For each finite-length power series $s(x) \;\; \pmod{x^{L_s}}$, do:
\begin{itemize}
\item	Apply the Berlekamp-Massey algorithm  to determine $\lambda(x)$.
\item   Compute $b(x)=s(x)\lambda(x) \pmod{x^{t-L_{xB}+1}}$.
\item	Construct $\Lambda'(x)=\lambda(x)\cdot \Lambda(x) + b(x)\cdot xB(x)$.
\item   Determine the distinct roots of $\Lambda'(x)$ within $\{\alpha^{-i}\}_{i=0}^{n-1}$.
\item   Compute error magnitudes using the Forney's formula if the number of distinct roots is equal to the degree $L_{\Lambda'}$. 
\end{itemize}

\item Return the list of codewords which are within distance $t$ from the received word.
\end{enumerate}

\begin{theorem}
Let a LECC $t$  satisfy \eqref{t-limit}. If  the multiplicity $m$ is chosen as in \eqref{m-bound} and subsequently the $y$-degree $P_y$ as in \eqref{y-deg},
then the proposed list decoding algorithm produces all codewords within distances $t$ from a received word.
\end{theorem}

\noindent
{\bf Remarks: } The proposed list decoding algorithm can be applied to 
the decoding of generalized Reed-Solomon codes, and error-and-erasure decoding using the method in \cite{Forney}.

Let $t_\text{opt}$ denote the achievable LECC
\begin{equation}
t_\text{opt}\eqdef \lceil n-1-\sqrt{n(n-d)} \rceil. \label{max-t}
\end{equation}
\begin{corollary}
The number of codewords that lie within (strictly less than) Hamming distance $n-\sqrt{n(n-d)}$ from a received word is $O(n-\sqrt{n(n-d)})$.
\end{corollary}
{\em Proof:} The number of such codewords is upper bounded by the degree $P_y$ that corresponds to $t=t_\text{opt}$.
Substituting \eqref{m-bound} into \eqref{y-deg}, we further obtain 
\begin{eqnarray*}
P_y&=& O\left( \frac{t_\text{opt}}{2t_\text{opt}-d} \cdot \frac{t_\text{opt}(d-t_\text{opt})}{t_\text{opt}^2-nt_\text{opt}+nd} \right) \\
&=& O\left( \frac{t_\text{opt}}{2t_\text{opt}-d} \cdot \frac{t_\text{opt}(d-t_\text{opt})} { (n-\sqrt{n(n-d)}- t_\text{opt})(n+\sqrt{n(n-d)}- t_\text{opt}) } \right)\\
&=& O\left( \frac{t_\text{opt}}{2t_\text{opt}-d} \cdot \frac{t_\text{opt}(2t_0-t_\text{opt})} { n+\sqrt{n(n-d)}- t_\text{opt} } \right)\\
&=& O\left( \frac{n-\sqrt{n(n-d)}}{2n-2\sqrt{n(n-d)}-d} \cdot \frac{(n-\sqrt{n(n-d)})(d-(n-\sqrt{n(n-d)}))} { n+\sqrt{n-d}- (n+\sqrt{n(n-d)}) } \right)\\
&=& O\left( n-\sqrt{n(n-d)} \right)  
\end{eqnarray*}
as desired. 

We next justify the treatment of  the term $n-\sqrt{n(n-d)}- t_\text{opt}$ as $O(1)$. Indeed, $n-\sqrt{n(n-d)}- t_\text{opt}$ 
can be arbitrarily small with appropriate choices of $n$.  However, if we replace $t_\text{opt}$ by 
$$t'=\lfloor n-\sqrt{n(n-d)}-0.5\rfloor,$$
then we immediately have 
$ 0.5 \leq n-\sqrt{n(n-d)}- t' <1.5,$
i.e., $n-\sqrt{n(n-d)}- t'=O(1)$. On the other hand, we have $0\leq t_\text{opt}-t'\leq 1,$  in which difference by 1 can be safely ignored 
in light of asymptotic behavior.     \hfill $\Box\Box$

\noindent
{\bf Remarks:} In \cite{Guru-Sudan}, the product term $(n-\sqrt{n(n-d)}- t_\text{opt})(n+\sqrt{n(n-d)}- t_\text{opt})$ is regarded as constant $O(1)$,
which results in a quadratic bound $O(\sqrt{kn^3})$ for list size, which otherwise is a linear term as well. 
Given a LECC $t<n-\sqrt{n(n-d)}$ for $(n,\; k)$ Reed-Solomon code, 
the explicit construction of the Guruswami-Sudan algorithm is given in \cite{Guru-Sudan}:
$$P_y=\bigg\lfloor \frac{1}{n-d}\bigg( (n-t)\big( 1+   
\big\lfloor \frac{n(n-d)+\sqrt{n^2(n-d)^2+4( (n-t)^2-n(n-d) )} }{ 2( (n-t)^2 -n(n-d) )} \big\rfloor
\big) -1     \bigg) \bigg\rfloor $$
whereas the explicit construction of the proposed algorithm shows
$$P_y=\bigg\lfloor \frac{t}{2t-2t_0} \cdot \bigg\lfloor \frac{t(2t_0-t)}{t^2-2n(t-t_0)} \bigg\rfloor \bigg\rfloor.$$
Figures~\ref{FIG-perf255} and \ref{FIG-perf2047} shed some light on the tightness of
the list size bound between the two algorithms.
Note that due to integer precision, 
it is possible that under a fixed multiplicity $m$, larger LECC $t$ causes smaller list size, specifically, let $t$ and $t'$, $t<t'$ 
be permissible for the same $m$, 
then it holds $\lfloor \frac{tm}{2(t-t_0)}\rfloor \geq \lfloor \frac{t'm}{2(t'-t_0)}\rfloor$. 
To this end, the minimum value should be chosen in the range where the same multiplicity $m$ applies. 

Ignoring the integer precision, the ratio $\frac{P_{y, \text{GS}}}{P_{y, \text{prop}}}$ with respect to the LECC limit
$t=n(1-\sqrt{1-D})$ is expressed as 
$$\frac{P_{y, \text{GS}}}{P_{y, \text{prop}}} \bigg|_{t=n(1-\sqrt{1-D})} = \frac{1}{1-\sqrt{1-D}},$$
which reveals that the derivative bound of the proposed algorithm is {\em universally} tighter than that of the Guruswami-Sudan algorithm 
on the boundary of the optimal LECC, as illustrated in Figure~\ref{FIG-perf-ratio}. 
Finally, we clarify that the above results are consistent with the work in \cite{Justesen, Ruck}, in which the list-$l$
Guruswami-Sudan bound  is shown to be tight only when the algorithm degenerates to the classical (list-1) hard-decision decoding.


\subsection{Complexity Analysis}

Finally we analyze the computational complexity in terms of field operations, 
assuming $n$ and $k$ are large enough and the field cardinality $q\leq 2^n$ 
(which is used to bound the factorization complexity, as discussed in \cite{Guru-Sudan}). 
Following the convention, we will fix the normalized minimum distance $D=d/n$ while letting $n$ (and $d$) go to infinity.
Our particular interest lies in the case of  the achievable LECC $t_\text{opt}$ which is defined in \eqref{max-t}.
In this case, the multiplicity chosen in \eqref{m-bound} simplifies to
\begin{eqnarray*}
m &=& O\left( \frac{t_\text{opt}(d-t_\text{opt})}{t_\text{opt}^2-2nt_\text{opt}+nd} \right)\\
&=& O\left( \frac{t_\text{opt}(d-t_\text{opt})}{(n-\sqrt{n(n-d)}-t_\text{opt})(n+\sqrt{n(n-d)}-t_\text{opt})} \right) \\
&=& O\left( \frac{t_\text{opt}(d-t_\text{opt})}{n+\sqrt{n(n-d)}-t_\text{opt}} \right) \\
&=& O\left( \frac{(n-\sqrt{n(n-d)})(d-(n-\sqrt{n(n-d)})}{2\sqrt{n(n-d)}} \right) \\
&=& O\left( \left(\sqrt{n}-\sqrt{n-d}\right)^2 \right).  \label{m-asymp}
\end{eqnarray*}

To the best of our knowledge, the up-to-date most efficient interpolation technique is through the Koetter updating procedure 
which exhibits complexity $O(N^2)$ with $N$ linear constraints \cite{Koetter, Ma}. 
To achieve optimal performance, the number of constraints is of the order $O(nm^2)=O\left(n(\sqrt{n}-\sqrt{k})^4\right)$.
Thus, we have the following characterization regarding the interpolation complexity.
\begin{lemma}
The rational interpolation of the proposed list decoding algorithm 
can be implemented with the complexity of 
$$O\left(n^2(\sqrt{n}-\sqrt{n-d})^8\right)$$ 
to achieve the LECC $t_\text{opt}=\lceil n-1-\sqrt{n(n-d)} \rceil. $
\end{lemma}

We next analyze the complexity of the proposed factorization procedure, following the development in \cite{Roth}. 
Note the length of $s(x)$, which is factorized and used in the Berlekamp-Massey algorithm to 
obtain $\lambda(x)$ and $b(x)$, is bounded by 
$$3t-2t_0-\deg(\lambda(x)) \leq 3t-2t_0-\deg(\lambda(x)) + (t-L_{xB})= 4(t-t_0).$$
In \cite{Lidl}, it is shown that the roots in $\GF(q)$ of a polynomial of degree $u$ can be found in expected time complexity 
$O(u^2 \cdot\log^2 u \cdot \log q)$. We observe that in each iteration ``$^{(i)}$" 
the degrees  of the polynomials $\mQ^{(i)}_j(0, y)$ satisfy $\sum_j \deg (\mQ^{(i)}_j(0, y)) \leq P_y$ (See Lemma~6.2 in \cite{Roth}). 
The root-finding complexity in one iteration is then bounded by
$$\sum_j O( \deg^2 (\mQ^{(i)}_j(0, y)) \cdot\log^2 \deg (\mQ^{(i)}_j(0, y)) \cdot \log q ) \leq O( P_y^2 \cdot\log^2 P_y \cdot \log q ).$$
Thus, the computational complexity of determining roots of $\mQ(0, y)$ in up to $4(t-t_0)$ iterations is bounded by 
$$O\left((t-t_0)\cdot P_y^2 \cdot \log^2 P_y^2 \cdot \log q \right).$$ 

We now analyze the shift operation $\hmQ^{(i)}(x, y)\gets \mQ^{(i)}(x, y+s_i)$.  
Lemma~6.2 in \cite{Roth} indicates that the step $\mQ^{(i+1)}(x, y)\gets \hmQ^{(i)}(x, xy)/x^a$, 
where $x^a$ denotes the largest power of $x$ which divides $\hmQ^{(i)}(x, xy)$, results in 
$$\deg \left( \mQ^{(i+1)}(0, y) \right) \leq a    \leq h    \leq    \deg \left( \mQ^{(i)}(0, y) \right),$$
where $h$ denotes the multiplicity of the root $s_i$ of  $\mQ^{(i)}(0, y)$.
Hence, the parameter $a$ is non-increasing with iterations going on. 
Therefore, it suffices, in each iteration, to update the first $4a(t-t_0)$ terms of $[y^l]\mQ^{(i)}(x, y)$, $l=0, 1, 2, \ldots, P_y$, 
in order to determine the first $4(t-t_0)$ terms of $s(x)$. Since each term can be updated with $O(P_y)$ operations, 
the corresponding complexity in one iteration is 
$$\sum_j 4a_j\cdot (t-t_0)P_y^2 \leq \sum_j 4 \deg ( \mQ^{(i)}_j(0, y) ) \cdot (t-t_0)P_y^2 =O\left( (t-t_0) P_y^3 \right).$$
and thus the resulting overall complexity is $O\left((t-t_0)^2 P_y^3\right)$. 

Finally, since there are at most $P_y$ candidate polynomials $s(x)$, it takes $O(P_y (t-t_0)^2)$ operations to  compute 
the corresponding $\lambda(x)$ and $b(x)$ through the Berlekamp-Massey algorithm.
Clearly, the overall complexity of the proposed factorization procedure is dominated by the shift operation with the complexity of
$$O\left( (t-t_0)^2 P_y^3 \right)=O\left( n^{3/2} (\sqrt{n}-\sqrt{n-d})^7 \right).$$

Therefore, we characterize the complexity of the factorization procedure as follows
\begin{lemma}
Given that the field cardinality is at most $2^n$, the rational factorization procedure of the proposed list decoding algorithm 
can be implemented with the complexity of
$$O\left(n^{3/2}(\sqrt{n}-\sqrt{n-d})^7\right)$$ 
to achieve the LECC $t_\text{opt}$.
\end{lemma}

The following theorem summarizes the complexity of the proposed algorithm. 
\begin{theorem}
Given that the field cardinality is at most $2^n$,
the proposed list decoding algorithm exhibits the computational complexity in terms of field operations
\begin{equation}
O\left(n^{2}(\sqrt{n}-\sqrt{n-d})^8\right)
\end{equation}
to achieve  the LECC $t_\text{opt}=\lceil n-1-\sqrt{n(n-d)} \rceil$.
\end{theorem}

\begin{figure*}[t] 
\centering
\includegraphics[width=6in]{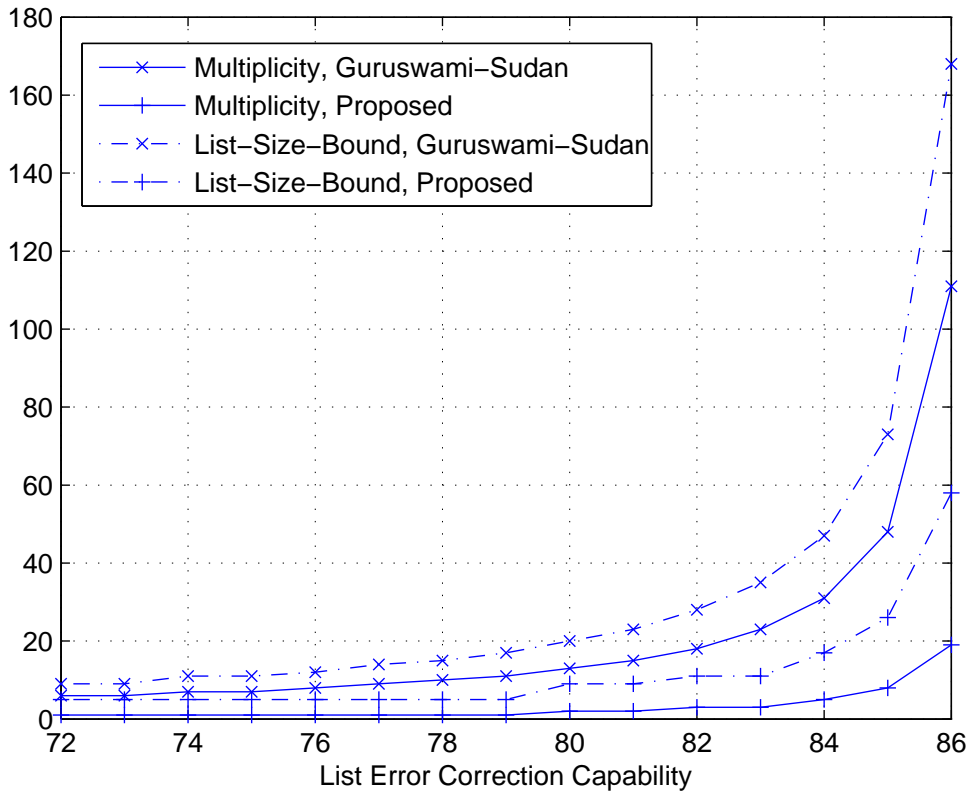}
\caption{Multiplicity and list-size-bound as functions of list error correction capability for the (255, 112) Reed-Solomon code. \label{FIG-perf255}}
\end{figure*}

\begin{figure*}[t] 
\centering
\includegraphics[width=6in]{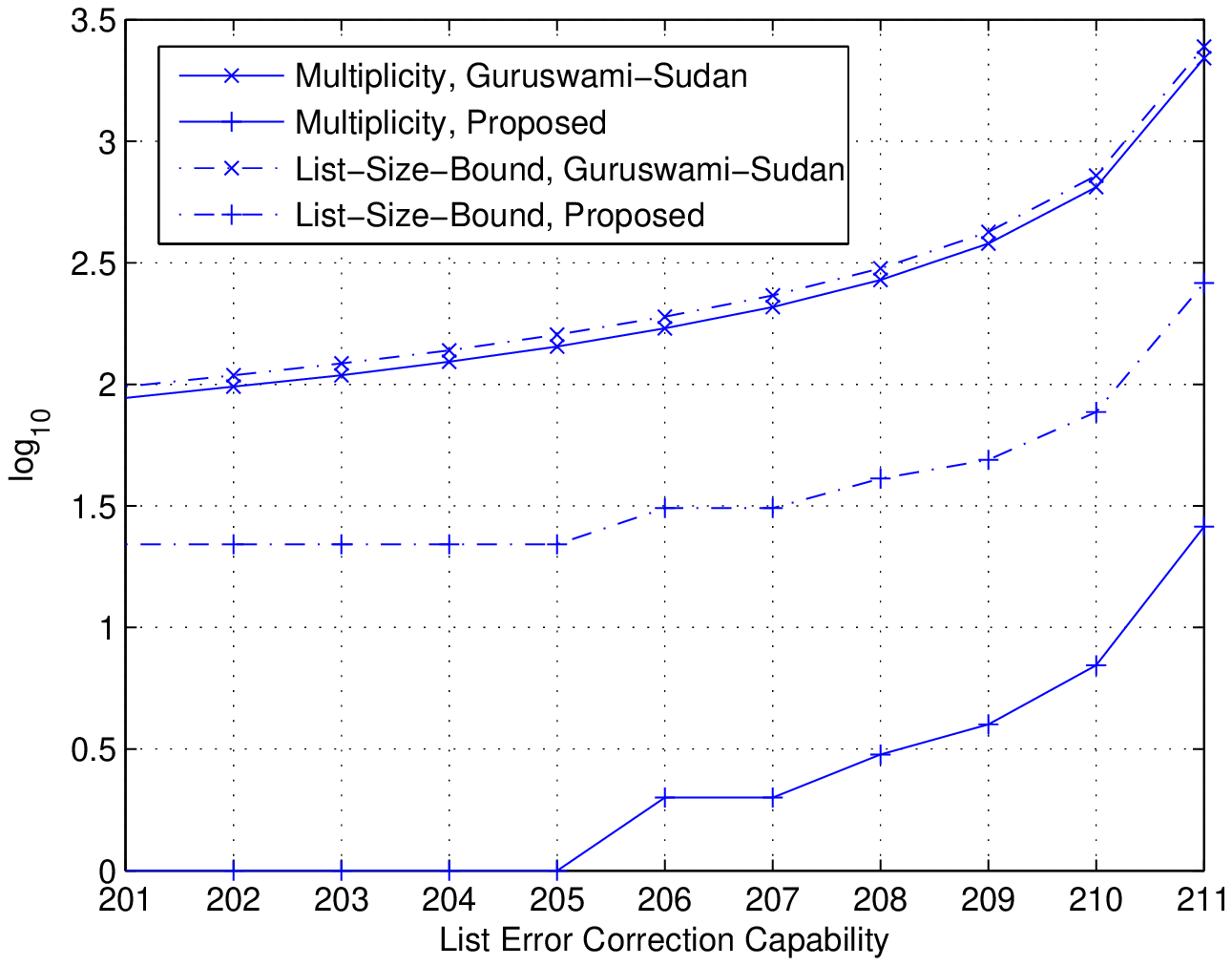}
\caption{Multiplicity and list-size-bound as functions of list error correction capability for the (2047, 1647) Reed-Solomon code. \label{FIG-perf2047}}
\end{figure*}

\noindent
{\bf Remarks: } 
Note that the multiplicity dictates the complexity. It is insightful to also compare the minimum multiplicities 
between the Guruswami-Sudan algorithm and the proposed algorithm. For an intermediate $\frac{n-k}{2} <t<n-\sqrt{n(n-d)}$,
 the value of multiplicity given in \cite{Guru-Sudan} is, under our notations, 
$$m_\text{GS}=1+\bigg\lfloor \frac{n(n-d)+\sqrt{(n^2(n-d)^2+4( (n-t)^2-n(n-d) )} }{ 2( t^2-2nt+nd )} \bigg\rfloor$$
whereas the value of multiplicity of the proposed approach is given in \eqref{m-bound}, i.e.,
$$ m_\text{prop}= 1+  \bigg\lfloor \frac{2(t-t_0)(n-t)}{t^2-2nt+nd}  \bigg\rfloor.$$
(recall that the above $m_\text{prop}$ is a sufficient value but not necessarily the minimum/optimal value, as indicated in \eqref{m-range}.)

Figures~\ref{FIG-perf255} and \ref{FIG-perf2047} plot the above two multiplicity functions for the (255, 112) Reed-Solomon code 
and the (2047, 1647) Reed-Solomon code, respectively. We observe in Figure~\ref{FIG-perf2047}, where the code rate is 0.8,
that to correct extra five erroneous symbols, the proposed algorithm requires multiplicity $m=1$, whereas the Guruswami-Sudan algorithm requires multiplicity 
$m=143$; to achieve the LECC limit of correcting extra 11 errors, the former requires the multiplicity $m=26$, whereas the latter requires $m=2197$.

\begin{figure*}[t] 
\centering
\includegraphics[width=6in]{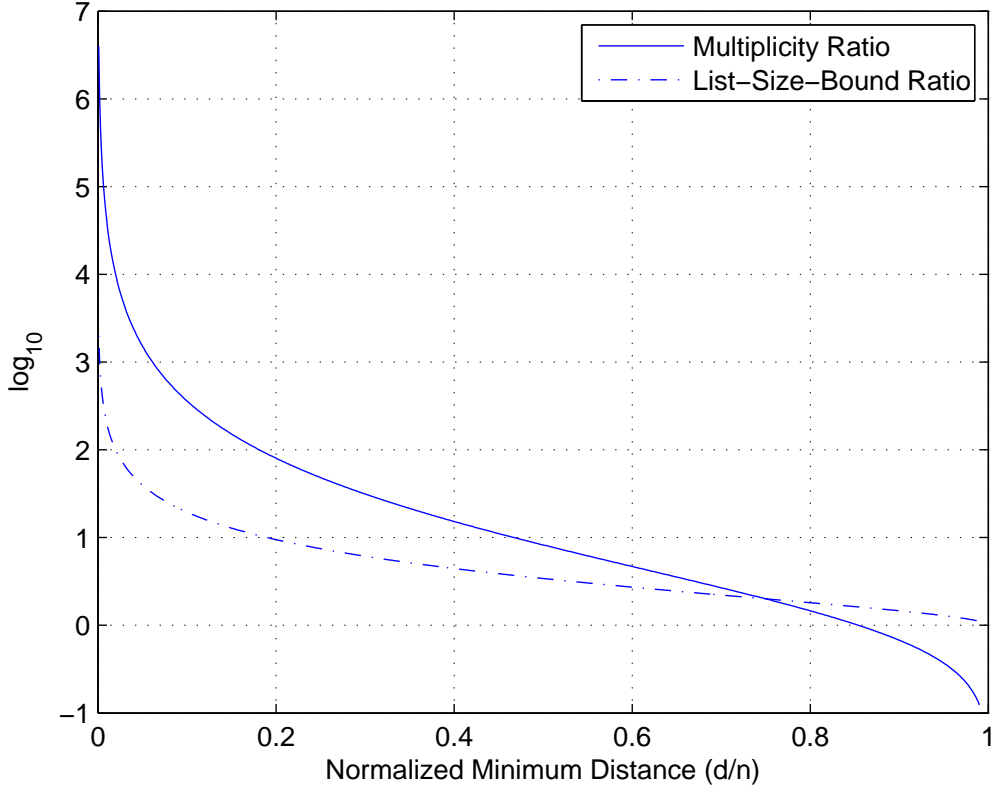}
\caption{The ratio of multiplicity and list-size-bound of the Guruswami-Sudan algorithm to that of the proposed algorithm 
in achieving the limit of list error correction capability. \label{FIG-perf-ratio}}
\end{figure*}

We further consider the ratio $\frac{m_\text{GS}}{m_\text{prop}}$ with respect to the LECC limit $t=n(1-\sqrt{1-D})$, which is expressed as 
$$\frac{m_\text{GS}}{m_\text{prop}}\bigg|_{t=n(1-\sqrt{1-D})} = \frac{n(n-d)}{2(t-t_0)(n-t)}= \frac{ \sqrt{1-D} }{ (1-\sqrt{1-D})^2 }.$$
(Note the integer precision is not taken into account in the above equality.)
It is plotted in Figure~\ref{FIG-perf-ratio} as a function of normalized minimum distance $D$. 
Evidently, the proposed algorithm reduces the (required) multiplicity (to achieve the optimal LECC) by orders of magnitude.

\subsection{An Alternative Perspective}

In the above we have characterized the multiplicity $m$ with respect to a fixed LECC $t$ as in \eqref{m-range}, and in particular, 
its asymptotics with respect to the maximum LECC $t_\text{opt}$ as in \eqref{m-asymp}.
In this subsection, we characterize the LECC with respect to a fixed multiplicity 
and show that the derivative algorithm has quadratic complexity which is identical to that of the Berlekamp-Massey algorithm. 
To simplify our analysis we will treat $P_y$, and subsequently $\deg_{1, w}(\mQ(x, y))$, as real numbers. Indeed, this treatment reveals many
fundamental insights.

Note that the degrees of freedom is maximized to 
$$ \max\{N_\text{free}\} = \frac{(tm+t-t_0)^2}{4(t-t_0)} $$
associated with $P_y=\frac{tm-t+t_0}{2(t-t_0)}$.
Consequently, the constraint $\max\{N_\text{free}\}>N_\text{cstr}$ with respect to a fixed multiplicity $m$ is reduced to 
\begin{equation}
\frac{(tm-t+t_0)^2}{4(t-t_0)}>\frac{nm(m+1)}{2},
\end{equation}
which simplifies to
$$ t^2(m+1)^2 -2t(m+1)(t_0+mn) +t_0^2+2t_0nm(m+1) >0.$$
Solving, we obtain 
\begin{equation}
t< \frac{1}{m+1} t_0 + \frac{m}{m+1}\left(n-\sqrt{n(n-d)} \right).  \label{t-m-linear}
\end{equation}
It indicates that, the practical LECC gain can be achieved with very small constant multiplicity, 
irregardless of the code length $n$, and particularly, half the LECC gain is achieved with multiplicity $m=1$. 

\begin{figure}[t] 
\centering
\includegraphics[width=6in]{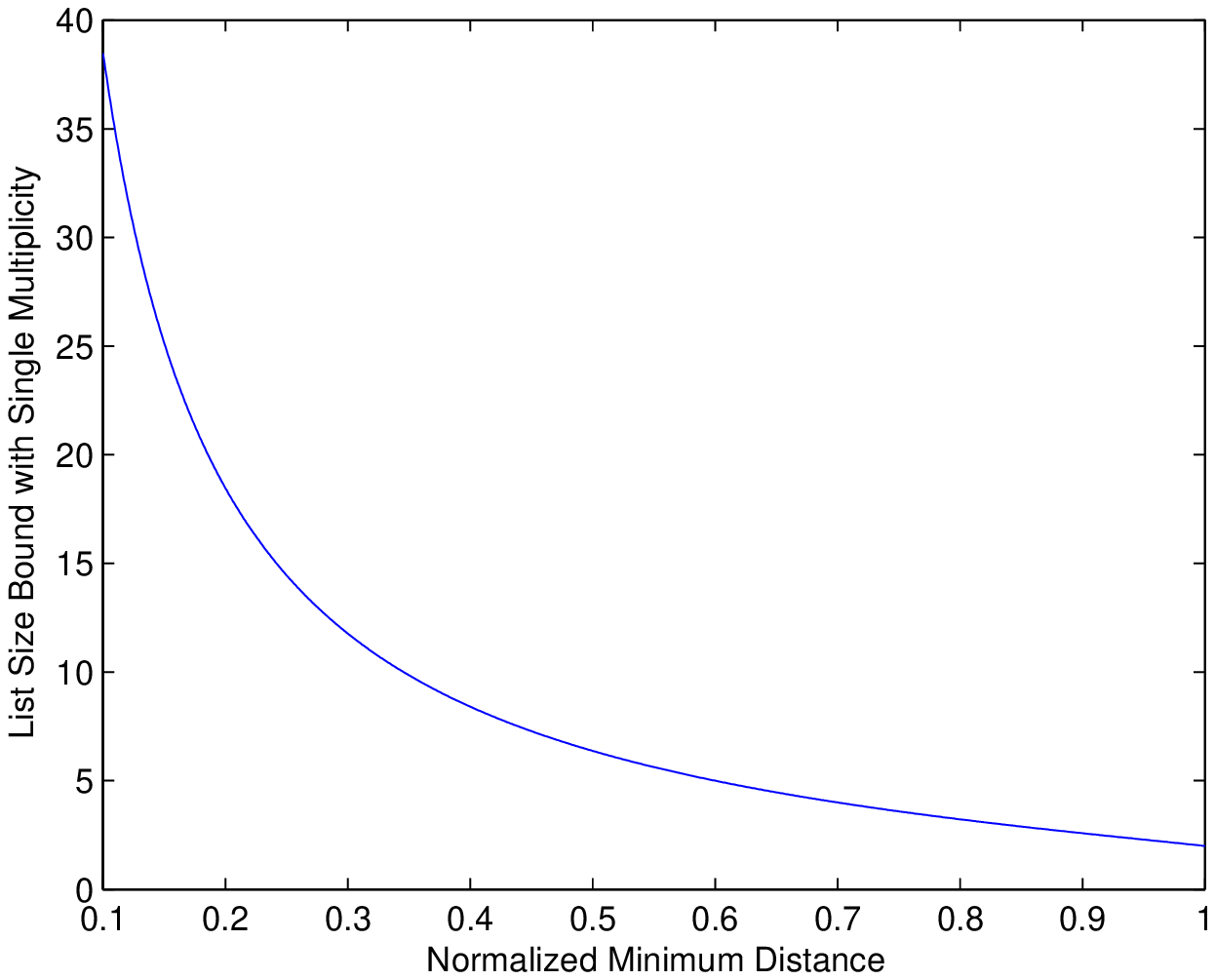}
\caption{The bound on the list size with multiplicity $m=1$. \label{FIG-listsize}}
\end{figure}

Recall that the above LECC is achieved with $P_y$ set to 
\begin{eqnarray}
P_y\aline{=}\frac{tm-t+t_0}{2(t-t_0)}  \nonumber \\
\aline{=} \frac{D}{(1-\sqrt{1-D})^2 } + \frac{m-1}{1-\sqrt{1-D}},
\end{eqnarray}
where the second equality is obtained by substituting $t$ with the limit in \eqref{t-m-linear}. 
It indicates that the list size is upper bounded by a constant for a fixed normalized minimum distance $D$, regardless of code length $n$. 
Figure~\ref{FIG-listsize} also reveals that the bound increases when the distance $D$ decreases, 
although the algorithm itself becomes less and less powerful.
On the other hand, the algorithmic complexity is quadratic following the step-by-step analysis of the preceding subsection,
fundamentally attributed to a constant upper bound on the list size. 

The above two-fold facts immediately reveals the following fundamental insights
\begin{theorem}
For arbitrarily small $\epsilon>0$, the list decoding up to the LECC 
\begin{equation}
t=\left\lfloor \epsilon \cdot t_0+ (1-\epsilon)\cdot (n-\sqrt{n(n-d)})  \right\rfloor
\end{equation}
 can be achieved by 
the proposed algorithm with multiplicity $m=\lfloor \frac{1}{\epsilon} \rfloor$, whose complexity is quadratic in nature, $O(n^2)$.  
\end{theorem}

\begin{figure}[t] 
\centering
\includegraphics[width=6.4in]{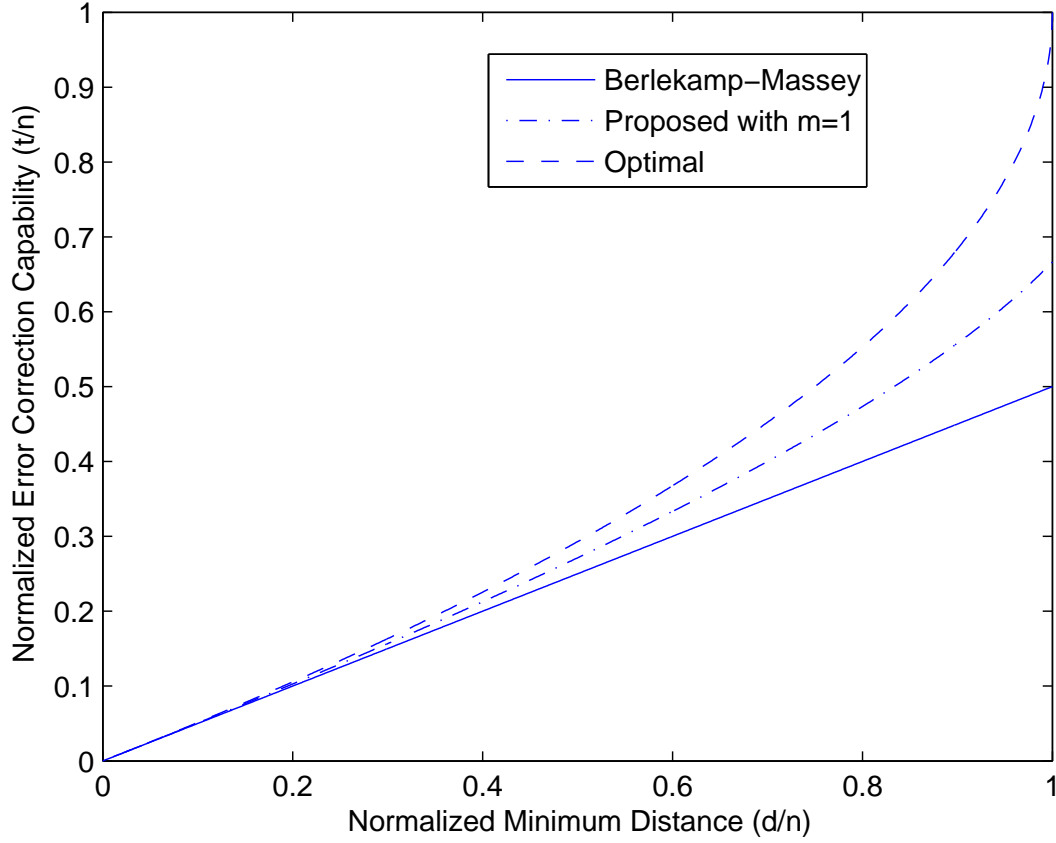}
\caption{Normalized (list) error correction capability as a function of the normalized minimum distance on decoding Reed-Solomon codes. 
\label{FIG-RS-bound}}
\end{figure}

\noindent
{\bf Remarks: } $(i)$. Even if $m=1$ the proposed algorithm universally outperforms the conventional hard-decision decoding,
 as shown in Figure~\ref{FIG-RS-bound}, 
in contrast to the Sudan algorithm, 
where the improvement is observed only when the code rate $\frac{k}{n}<\frac{1}{3}$ \cite{Sudan}. 
$(ii)$. The list size of candidate codewords is upper bounded by a constant 
arbitrarily close to the LECC limit $n-\sqrt{n(n-d)}$. 
However, the list size may be as large as superpolynomial slightly beyond the LECC limit, as explicitly constructed in \cite{Guru}.


\section{\sc Improved List Decoding Algorithm for Binary BCH Codes}

In this section we investigate the list decoding of the narrow-sense binary BCH codes. 
It is straightforward to apply the algorithm in the preceding section to the decoding of a  binary BCH code 
and obtain the inherent LECC to be $n(1-\sqrt{1-D})$, 
where $D$ denotes the normalized designed minimum distance. 
In the following we present an improved list decoding algorithm for the  binary BCH codes that achieves the Johnson bound
for the binary codes $\frac{n}{2}(1-\sqrt{1-2D})$, which gives a general lower bound on the number of correctable errors using 
small lists for any code, as a function of the normalized minimum distance of the code  \cite{Johnson}.

The following lemma identifies a special feature of binary BCH codes. 
Its proof follows from Lemma~\ref{LEM-Lambda-form} and the particularity of the Berlekamp algorithm.
\begin{lemma}   \label{LEM-Lambda-BCH-form}
Let $\Lambda^*(x)$ be the true error locator polynomial as defined in \eqref{def-true-Lambda}. 
Let $\Lambda(x)$ and $B(x)$ be the error locator and correction polynomials, respectively, 
obtained from the (re-formulated) Berlekamp algorithm. Then, 
$\Lambda^*(x)$ exhibits the form of
\begin{equation}
\Lambda^*(x)= \Lambda(x) \cdot \lambda^*(x^2) + x^2B(x) \cdot b^*(x^2), \label{form-Lambda-BCH}
\end{equation}
where the polynomials $\lambda^*(x)$ and $b^*(x)$ exhibit the following properties:\\
$(i)$. $\lambda^*_0=1$; \\
$(ii)$. if $b^*(x) = 0$, then $\lambda^*(x)= 1$; \\
$(iii)$. $\lambda^*(x)$ and $b^*(x)$ are coprime;\\
$(iv)$. $2\deg(\lambda^*(x))= \deg(\Lambda^*(x))-L_\Lambda$ and $2\deg(b^*(x))\leq \deg(\Lambda^*(x))-L_{x^2B}$, or \\ 
 $2\deg(\lambda^*(x))\leq \deg(\Lambda^*(x))-L_\Lambda$ and $2\deg(b^*(x)) = \deg(\Lambda^*(x))-L_{x^2B}$;\\
$(v)$. if $\deg(\Lambda^*(x))<d$, then $\lambda^*(x)$ and $b^*(x)$ are unique;
\end{lemma} 

We proceed to incorporate the special form \eqref{form-Lambda-BCH} into the rational interpolation process 
to optimize the LECC.
Define
\begin{equation}
y_i=-\frac{\Lambda(\alpha^{-i})}{\alpha^{-2i} B(\alpha^{-i})}, \hspace{0.2in} i=0, 1, 2, \ldots, n-1. \label{def-y-value-BCH}
\end{equation} 
Note we are interested in determining the form of $y\lambda(x^2)-b(x^2)$, so we naturally assign the weight of $y$ to be
\begin{equation}
w=L_\Lambda-L_{x^2B},  \label{def-w-BCH}
\end{equation}
which is always odd since $L_\Lambda+L_{x^2B}=d$ is odd.

\begin{lemma}
Let $\Lambda(x)$ and $B(x)$ be the error locator and correction polynomials, respectively, obtained from the (re-formulated) Berlekamp algorithm.
Let $\mQ(x, y)$ be a bivariate polynomial passing through all $n$ points
$\{ (1, y_0)$, $(\alpha^{-2}, y_1)$,  $(\alpha^{-4}, y_2)$, \ldots, $(\alpha^{-2(n-1)}, y_{n-1}) \}$ (where $y_i$ is defined in \eqref{def-y-value-BCH}),
each with multiplicity $m$. If
\begin{equation}
(t-L_\Lambda)P_y + \deg_{2, w}(\mQ(x, y))  < 2tm,  \label{zero-constraint-BCH}
\end{equation}
where $P_y$ denotes the power of $y$ in $\mQ(x, y)$ and $\deg_{2, w}(\mQ(x, y))$ (where $w$ is defined in \eqref{def-w-BCH})
 denotes the $(2, w)$-weighted degree of $\mQ(x, y)$, then $\mQ(x^2, y)$ contains all factors of the form $y\lambda(x^2)-b(x^2)$ 
which pass through $t$ ($t\geq L_\Lambda$) points.
\end{lemma}
{\em Proof: } Let $y\lambda(x^2) - b(x^2)$ be a polynomial passing through $t$ out of $n$ points, 
$\{ (1, y_0)$, $(\alpha^{-1}, y_1)$,  $(\alpha^{-2}, y_2)$, \ldots, $(\alpha^{-(n-1)}, y_{n-1}) \}$, 
and $(\alpha^{-i}, y_i)$ be one of points.
Let 
$$p(x)\eqdef \frac{b((x+\alpha^{-i})^2)}{\lambda((x+\alpha^{-i})^2)}-y_i
=\frac{b(x^2+\alpha^{-2i})}{\lambda(x^2+\alpha^{-2i})}-y_i,$$ 
where the second equality is due to the field of characteristic of 2.
Then $p(0)=0$ and moreover $p(x)$ must be in the form of
$$p(x)=x^2p'(x),$$
where $p'(x)$ contains no pole of zero.
We next show that $(x-\alpha^{-i})^{2m}$ divides the following polynomial 
$$g(x)\eqdef \lambda^{P_y}(x^2) \mQ\left(x^2, \;\; \frac{b(x^2)}{\lambda(x^2)}\right)$$
To this end, consider 
$$g'(x)\eqdef \lambda^{P_y}(x^2+\alpha^{-2i})\cdot \mQ^{(i)}(x^2, p(x)),$$
where $\mQ^{(i)}(x, y)$ is the shift of $\mQ(x, y)$ to $(\alpha^{-2i}, y_i)$.
Consequently, 
\begin{eqnarray*}
g(x)&=&\lambda^{P_y}(x^2)\mQ^{(i)}\left(x^2+\alpha^{-2i}, \;\; \frac{b(x^2)}{\lambda(x^2)}-y_i\right) \\
&=& \lambda^{P_y}(x^2)\mQ^{(i)}\left(x^2+\alpha^{-2i}, \;\; p(x-\alpha^{-i})\right) = g'(x-\alpha^{-i}).
\end{eqnarray*}
By construction, $\mQ^{(i)}(x, y)$ has weighted degree less than $m$. Thus, plugging in $p(x)=x^2p'(x)$,
we conclude that $(x-\alpha^{-i})^{2m}$ divides $g(x)$.
Therefore, $g(x)$ is a polynomial that contains at least $2tm$ roots, i.e., all roots of $\Lambda^*(x)$ 
each with multiplicity $2m$. The remaining proof trivially follows that of Lemma~\ref{LEM-zero-constraint}. \hfill $\Box\Box$

Note that the degrees of freedom in $\mQ(x, y)$, given the degrees $\deg_{2, w}(\mQ(x, y))$ and $P_y$, is 
\begin{eqnarray*}
\sum_{i=0}^{P_y} \left(1+\lfloor\frac{\deg_{2, w}(\mQ(x, y))-iw}{2}\rfloor \right) 
\aline{\geq} \sum_{i=0}^{P_y} \left(1+\frac{\deg_{2, w}(\mQ(x, y))-iw-1}{2}\right)   \\
\aline{=} \frac{(\deg_{2, w}(\mQ(x, y))+1-P_yw/2)(P_y+1)}{2}.
\end{eqnarray*}
Define
$$t_0\eqdef L_\Lambda-w/2=\frac{L_\Lambda+L_{x^2B}}{2}=\frac{d}{2}.$$
Note that the above definition is consistent with the case of Reed-Solomon codes.

We next maximize the lower bound of $N_\text{free}$, $\frac{(\deg_{2, w}(\mQ(x, y))+1-P_yw/2)(P_y+1)}{2}$, subject to fixed number of errors $t$, 
fixed multiplicity $m$ and the zero constraint \eqref{zero-constraint-BCH}
\begin{eqnarray}
\frac{(\deg_{2, w}(\mQ(x, y))+1-P_yw/2)(P_y+1)}{2} &\leq& \frac{1}{2} (2tm-P_y(t-t_0))(P_y+1) \nonumber \\
&=& -\frac{t-t_0}{2} P_y^2 + \frac{2tm-t+t_0}{2} P_y + tm \nonumber \\
&=& -\frac{t-t_0}{2}\left(P_y-\frac{2tm-t+t_0}{2(t-t_0)}\right)^2 + \frac{(2tm+t-t_0)^2}{8(t-t_0)},  \label{max-freedom-BCH}
\end{eqnarray}
where ``=" in the ``$\leq$" is  achieved if and only if
\begin{equation}
\deg_{2, w}(\mQ^*(x, y))=2tm-1-P_y(t-L_\Lambda). \label{Q-deg-BCH}
\end{equation}

We note that the maximum degrees of freedom is achieved by choosing $P_y^*$ to be the closest integer to $\frac{2tm-t+t_0}{2t-2t_0}$, 
 i.e.,
\begin{equation}
P_y^*=\bigg\lfloor \frac{2tm-t+t_0}{2t-2t_0} +0.5 \bigg\rfloor = \bigg\lfloor \frac{tm}{t-t_0}\bigg\rfloor . \label{y-deg-BCH}
\end{equation}
Therefore, the maximum degrees of freedom has the following lower bound
\begin{equation}
 -\frac{t-t_0}{8} +  \frac{ ( 2tm+t-t_0 )^2 }{ 8(t-t_0) }= \frac{ t^2m^2+tm(t-t_0) }{ 2(t-t_0) }. 
\end{equation}
Hence, to solve for the linear equation system, it suffices to enforce
\begin{equation}
\frac{ t^2m^2+tm(t-t_0) }{ 2(t-t_0) } > \frac{nm(m+1)}{2},
\end{equation}
which is equivalent to
$$m^2(t^2-n(t-t_0))-m(t-t_0)(n-t) > 0.$$
The above condition holds true for sufficiently large $m$ if and only if 
$$t^2-n(t-t_0)>0,$$
which indicates the following limit of LECC
\begin{equation}
t<\frac{n-\sqrt{n(n-4t_0)} }{2} = \frac{n-\sqrt{n(n-2d)} }{2}.  \label{t-bound-BCH}
\end{equation}
It is always superior to the inherent LECC $n-\sqrt{n(n-d)}$, following the fact
$$\frac{n-\sqrt{n(n-2d)}}{2}-\left(n-\sqrt{n(n-d)}\right)=\frac{t_0\sqrt{n}}{ \sqrt{n-d}+\sqrt{n-2d} } - 
\frac{t_0\sqrt{n}}{ \sqrt{n}+\sqrt{n-d} }>0.$$

When \eqref{t-bound-BCH} is satisfied, we may choose the multiplicity $m$ to be
\begin{equation}
m^* = \bigg\lfloor \frac{(t-t_0)(n-t)}{t^2-n(t-t_0)} +1 \bigg\rfloor =\bigg\lfloor \frac{tt_0}{t^2-n(t-t_0)} \bigg\rfloor.  \label{m-bound-BCH}
\end{equation}

The following presents the ``lossless" maximization of the degrees of freedom conditioned on the fixed $t$ and $m$. 
It can be easily verified that the degrees of freedom is bounded by
\begin{eqnarray*}
 \frac{(\deg_{2, w}(\mQ(x, y))+1-P_yw/2)(P_y+1)}{2} + \frac{P_y}{4}+\frac{1}{2} 
\aline{\geq} \sum_{i=0}^{P_y} \left(1+\lfloor\frac{\deg_{2, w}(\mQ(x, y))-iw}{2}\rfloor \right) \\
\aline{\geq}  \frac{(\deg_{2, w}(\mQ(x, y))+1-P_yw/2)(P_y+1)}{2} + \frac{P_y}{4}.
\end{eqnarray*} 
Due to the integer constraint, it suffices to treat and optimize the degrees of freedom as follows
\begin{eqnarray}
N_\text{free}\aline{=}\frac{(\deg_{2, w}(\mQ(x, y))+1-P_yw/2)(P_y+1)}{2} + \frac{P_y}{4} \nonumber \\
\aline{\leq} \frac{1}{2} (2tm-P_y(t-t_0))(P_y+1) +\frac{P_y}{4} \nonumber \\
\aline{=} -\frac{t-t_0}{2}\left(P_y-\frac{2tm-t+t_0+0.5}{2(t-t_0)}\right)^2 + \frac{(2tm-t+t_0+0.5)^2}{8(t-t_0)}+tm
\end{eqnarray}
where ``=" is achieved by choosing $\deg_{2, w}(\mQ(x, y))$ as in \eqref{Q-deg-BCH}.
Evidently, the optimal choice of $P_y$ is as follows, 
\begin{equation}
P_y^*=\left\lfloor \frac{2tm-t+t_0+0.5}{2(t-t_0)} +0.5 \right\rfloor=\left\lfloor \frac{tm+0.25}{t-t_0} \right\rfloor, \label{y-deg-BCH-new}
\end{equation}
which is slightly different from \eqref{y-deg-BCH}.
Consequently, the optimal multiplicity $m$ is the minimum one that complies with the following constraint
\begin{equation}
-\frac{t-t_0}{2}\left(\left\lfloor \frac{tm+0.25}{t-t_0} \right\rfloor -\frac{2tm-t+t_0+0.5}{2(t-t_0)}\right)^2 + \frac{(2tm-t+t_0+0.5)^2}{8(t-t_0)}+tm
>\frac{nm(m+1)}{2}. \label{org-free-cstr-BCH}
\end{equation}

We are ready to present the complete list decoding algorithm for binary BCH codes as follows.

{\large \bf \underline {List Decoding Algorithm for Binary BCH Codes}}
\begin{enumerate} 
\item[0.] Initialization: Input code length $n$, minimum distance $d$, and LECC $t$ satisfying 
$t<\frac{1}{2}\left(n-\sqrt{n(n-2d)} \right)$. 
Initialize the multiplicity $m$ based on \eqref{m-bound-BCH}, then greedily optimize it subject to the constraint \eqref{org-free-cstr-BCH},
 subsequently choose the $y$-degree $P_y$ as in \eqref{y-deg-BCH-new}. 

\item Input the received word and compute syndromes.

\item Apply the Berlekamp algorithm to determine $\Lambda(x)$ and $B(x)$. If $L_\Lambda >t$ then declare a decoding  failure.

\item Perform error correction and evaluate $y_i=-\frac{\Lambda(\alpha^{-i})} { \alpha^{-2i}B(\alpha^{-i}) } $, $i=0, 1, 2, \ldots, n-1$.

\item If $L_{x^2B}>t$,  then return the corresponding unique codeword when $\Lambda(x)$ has precisely $L_\Lambda$ valid roots, otherwise declare a decoding  failure.

\item Apply the rational interpolation procedure to compute a (2, $L_\Lambda-L_{x^2B}$)-weighted-degree polynomial $\mQ(x, y)$ 
that passes through $\left\{(\alpha^{-2i}, \; y_i)\right\}_{i=0}^{n-1}$, each with multiplicity $m$.

\item Apply the rational factorization process to obtain finite-length power series $s(x)\;\; \pmod{x^{L_s}}$ of the rational functions 
$\frac{b(x)}{\lambda(x)}$.

\item For each finite-length power series $s(x) \;\; \pmod{x^{L_s}}$, do:
\begin{itemize}
\item	Apply the Berlekamp-Massey algorithm  to determine $\lambda(x)$.
\item   Compute $b(x)=s(x)\lambda(x) \pmod{x^{t-L_{x^2B}+1}}$.
\item	Construct $\Lambda'(x)=\lambda(x^2)\cdot \Lambda(x) + b(x^2)\cdot x^2B(x)$.
\item   Determine the distinct roots of $\Lambda'(x)$ within $\{\alpha^{-i}\}_{i=0}^{n-1}$.
\end{itemize}

\item Return the list of codewords which are within distance $t$ from the received word.
\end{enumerate}

We summarize the above discussions into the following theorem
\begin{theorem}
For a given $(n, k, d)$ BCH code, let the LECC $t$ satisfies \eqref{t-bound-BCH} 
and the multiplicity $m$ be chosen as in \eqref{m-bound-BCH}, 
then the proposed list decoding algorithm produces all codewords 
up to distance $t$ from a received word.
\end{theorem}

{\bf Remarks:} The LECC limits of the proposed algorithm and the Guruswami-Sudan algorithm are illustrated in Figure~\ref{FIG-BCH-bound}. 
The proposed performance matches the Johnson bound for binary codes \cite{Johnson}. The proposed performance also demonstrates the tightness of 
the improved bound of the list size in \cite{Elias}.

Following the step-by-step complexity analysis in the preceding section, we characterize the complexity of the proposed algorithm 
 as follows
\begin{theorem}
Given that the field cardinality is at most $2^n$,
the proposed list decoding algorithm for $(n, k, d)$ binary BCH codes exhibits the computational complexity in terms of field operations
\begin{equation}
O\left(n^{2}(\sqrt{n}-\sqrt{n-2d})^8\right)
\end{equation}
to achieve the maximum LECC
$t_\text{opt} = \lceil \frac{n-\sqrt{n(n-2d)}}{2} -1 \rceil $.  
\end{theorem}

We proceed to characterize the proposed list decoding algorithm with respect to a fixed multiplicity $m$. For the conciseness of analysis,
 we treat the degrees of freedom to be $\frac{1}{2}(\deg_{2, w}(\mQ(x, y))+1-P_yw/2)(P_y+1)$, and $P_y$ as well as $\deg_{2, w}(\mQ(x, y))$ to be real numbers.
Following \eqref{max-freedom-BCH}, the maximum degrees of freedom is
$$\max\{N_\text{free}\}=\frac{(2tm+t-t_0)^2}{8(t-t_0)}$$
which is achieved by choosing $P_y=\frac{2tm-t+t_0}{2(t-t_0)}$.
Subsequently, the constraint of more degrees of freedom than the number linear constraints is reduced to
$$\frac{(2tm+t-t_0)^2}{8(t-t_0)}> \frac{nm(m+1)}{2}$$
which is re-organized in decreasing order of $t$ 
$$t^2(2m+1)^2 -2t((2m+1)t_0+2m(m+1)n) +t_0^2 +4t_0(m+1)mn >0.$$
Solving, we obtain the following LECC limit with respect to the fixed $m$
\begin{equation}
t<\frac{(2m+1)t_0 + 2m(m+1)n - 2m\sqrt{n^2(m+1)^2-nd(m+1)(2m+1)} }  {(2m+1)^2}.
\end{equation}
It can be shown that 
\begin{eqnarray}
\aline{}\frac{(2m+1)t_0 + 2m(m+1)n - 2m\sqrt{n^2(m+1)^2-nd(m+1)(2m+1)} }  {(2m+1)^2}  \nonumber \\
\aline{>} \frac{1}{2m+1} \cdot t_0 +\frac{2m}{2m+1} \cdot \frac{n-\sqrt{n(n-2d)}}{2}.
\end{eqnarray}
On the other hand, we have
\begin{eqnarray}
P_y\aline{=} \frac{2tm-t+t_0}{2(t-t_0)}  \nonumber \\
\aline{\leq} \frac{ (2m-1)\left(\frac{1}{2m+1} t_0 +\frac{2m}{2m+1} \frac{n-\sqrt{n(n-2d)}}{2} \right) +t_0 }
{ \frac{2}{2m+1} t_0 +\frac{4m}{2m+1} \frac{n-\sqrt{n(n-2d)}}{2} -d } \nonumber \\
\aline{=} \frac{2D}{ (1-\sqrt{1-2D})^2 } + \frac{2m-1}{1-\sqrt{1-2D}},
\end{eqnarray}
which indicates that the list size is upper bounded by a constant with respect to a given normalized minimum distance $D$. 
We characterize the above discussions into the following theorem.
\begin{theorem}
For arbitrarily small $\epsilon>0$, the list decoding of the $(n, k, d)$ binary BCH code up to the LECC 
\begin{equation}
t=\left\lfloor \epsilon \cdot t_0+ (1-\epsilon)\cdot \frac{n-\sqrt{n(n-2d)}}{2} \right\rfloor
\end{equation}
can be achieved by 
the proposed algorithm with multiplicity $m=\lfloor \frac{1}{2\epsilon} \rfloor$, whose complexity is quadratic in nature, $O(n^2)$.  
\end{theorem}

{\bf Example 6.} Consider the (63, 18, 21) BCH code. Its conventional error correction capability is $10$. 
The achievable LECC of the Guruswami-Sudan algorithm is $\lceil 63-\sqrt{63\times 42}-1 \rceil =11$,
whereas that of the proposed algorithm is $\lceil \frac{63-\sqrt{63\times 21}}{2}-1 \rceil=13$. 
To correct up to $t=13$ errors, the multiplicity and the degree of $y$ are set as follows. 
Let the multiplicity be $m=\lfloor \frac{13\times 10.5}{13^2-63\times(13-11.5)} \rfloor= 11$, following \eqref{m-bound-BCH}, 
which  turns out to be optimal. Let the $y$ degree be $P_y=\lfloor \frac{13\times 11+0.25}{13-10.5} \rfloor =57$, 
following \eqref{y-deg-BCH}, which also turns out to be optimal.
We next walk through a simulation example to illustrate the algorithmic procedure. Let the prototype codeword be 
\begin{eqnarray*}
\bc \aline{=} [\;1, 1, 0, 0, 0, 1, 0, 1, 0, 0, 1, 0, 1, 0, 1, 1, 0, 0, 1, 1, 0, 1, 1, 0, 1, 0, 0,  1, 0, 1, 1, \\
	\aline{}    0, 1, 0, 0, 1, 0, 1, 1, 0, 0, 0, 1, 1, 0, 1, 1, 1, 0, 1, 0, 0, 0, 0, 0, 0, 0, 1, 1, 0, 1, 1, 0\;]
\end{eqnarray*}
and the received word  be 
\begin{eqnarray*}
{\bf r} \aline{=} [\; 1, 1, 0, 0, 1, 1, 0, 0, 0, 0, 1, 0, 1, 0, 1, 1, 1, 1, 1, 0, 1, 1, 1, 0, 1, 0, 0, 1, 0, 1, 1, \\
		\aline{}   0, 1, 1, 0, 1, 0, 1, 1, 0, 0, 0, 1, 1, 0, 0, 0, 0, 0, 1, 1, 0, 0, 1, 0, 0, 0, 1, 1, 0, 0, 1, 0 \;]
 \end{eqnarray*}
which has 13 erroneous bits.
The true error locator polynomial is 
\begin{eqnarray*}
\Lambda^*(x) \aline{=} (1-\alpha^{4}x)(1-\alpha^{7}x)(1-\alpha^{16}x)(1-\alpha^{17}x)(1-\alpha^{19}x)(1-\alpha^{20}x)(1-\alpha^{33}x)(1-\alpha^{45}x)\\
\aline{} (1-\alpha^{46}x)(1-\alpha^{47}x) (1-\alpha^{50}x) (1-\alpha^{53}x) (1-\alpha^{60}x)  \\
\aline{=}  1 +\alpha^{29} x +\alpha^{23} x^{2}+\alpha^{37} x^{3}+\alpha^{61} x^{4}+\alpha^{42} x^{5}+\alpha^{28} x^{6}+\alpha^{15} x^{7}+\alpha^{35} x^{8}+\alpha^{40} x^{9}\\
\aline{}  +\alpha^{35} x^{10}+\alpha  x^{11}+\alpha^{12} x^{12}+\alpha^{39} x^{13} .
\end{eqnarray*}
The Berlekamp algorithm outputs the following pair of error locator and correction polynomials
\begin{eqnarray*}
\Lambda(x)\aline{=}1 + \alpha^{29} x  + \alpha^{16} x^{2} + \alpha^{5} x^{3} + \alpha^{2} x^{4} + \alpha^{47} x^{5} +
 \alpha^{6} x^{6} + \alpha^{49} x^{7} + \alpha^{27} x^{8} + \alpha^{19} x^{9} + \alpha^{34} x^{10} \\ 
B(x)\aline{=} \alpha^{2} + \alpha^{31} x  + \alpha^{23} x^{2} + \alpha^{50} x^{3} + \alpha^{51} x^{4} + \alpha^{42} x^{5}
 + \alpha^{53} x^{6} + \alpha^{53} x^{7} + \alpha^{32} x^{8} + \alpha^{15} x^{9}.
\end{eqnarray*}
In this case, the weight of $y$ is set to $w=L_\Lambda-L_{x^2B}=-1$.  
The following lists the $n=63$ interpolation points 
$\{ (\alpha^{-2i},\; \frac{\Lambda(\alpha^{-i})}{\alpha^{-2i}B(\alpha^{-i})} ) \}_{i=0}^{62}$:
{\small $$\begin{array}{lllllllll}
  (1,\; \alpha^{57})      & (\alpha^{-2},\; \alpha^{19})  &   (\alpha^{-4},\; \alpha^{54})  &   (\alpha^{-6},\; \alpha^{35})  &  
 (\alpha^{-8},\; \alpha^{13})  &   (\alpha^{-10},\; \alpha^{38})  &   (\alpha^{-12},\; \alpha^{48})  &   (\alpha^{-14},\; \alpha^{46})   \\
  
 (\alpha^{-16},\; \alpha^{27}) & (\alpha^{-18},\; \alpha^{49})  &   (\alpha^{-20},\; \alpha^{52})  &   (\alpha^{-22},\; \alpha^{58})  &    
  (\alpha^{-24},\; \alpha^{14})  &   (\alpha^{-26},\; \alpha^{34})  & (\alpha^{-28},\; \alpha^{45})  &   (\alpha^{-30},\; \alpha^{31})  \\
 
  (\alpha^{-32},\; \alpha^{5})  &   (\alpha^{-34},\; \alpha^{44}) & (\alpha^{-36},\; \alpha^{45})  &   (\alpha^{-38},\; \alpha^{3})  &    
 (\alpha^{-40},\; \alpha^{15})  &   (\alpha^{-42},\; \alpha^{61})  &   (\alpha^{-44},\; \alpha^{62})  &  (\alpha^{-46},\; \alpha^{30})  \\
 
	(\alpha^{-48},\; \alpha^{31})  &   (\alpha^{-50},\; \infty)  &   (\alpha^{-52},\; \alpha^{49})  & (\alpha^{-54},\; \alpha^{8})  & 
  (\alpha^{-56},\; \alpha^{35})  &   (\alpha^{-58},\; \alpha^{41})  &   (\alpha^{-60},\; \alpha^{57})  &   (\alpha^{-62},\; \alpha^{4})  \\
  
   (\alpha^{-1},\; \alpha^{45})  &   (\alpha^{-3},\; \alpha^{17})  &   (\alpha^{-5},\; \alpha^{34})  &   (\alpha^{-7},\; \alpha^{5})  & 
   (\alpha^{-9},\; \alpha^{16})  &   (\alpha^{-11},\; \alpha^{62})  &   (\alpha^{-13},\; \alpha^{7})  &   (\alpha^{-15},\; \alpha^{47})  \\
   
   (\alpha^{-17},\; \alpha^{26})  &   (\alpha^{-19},\; \alpha^{8})  &   (\alpha^{-21},\; \infty)  &   (\alpha^{-23},\; \alpha^{12})  &  
 (\alpha^{-25},\; \alpha^{54}) &      (\alpha^{-27},\;\alpha^{38})  &   (\alpha^{-29},\; \alpha^{11})  &   (\alpha^{-31},\; \alpha^{25})  \\

   (\alpha^{-33},\; \alpha^{62})  &   (\alpha^{-35},\; \alpha^{8})  &  (\alpha^{-37},\; \alpha^{59})  &   (\alpha^{-39},\; \alpha^{8})  &  
   (\alpha^{-41},\; \alpha^{30})  &   (\alpha^{-43},\; \alpha^{53}) & (\alpha^{-45},\; \alpha^{12})  &   (\alpha^{-47},\; \alpha^{47})  \\

  (\alpha^{-49},\; \alpha^{44})  &   (\alpha^{-51},\; \alpha^{15})  &   (\alpha^{-53},\; \alpha^{49})  &   (\alpha^{-55},\; \alpha^{26})  &   
	(\alpha^{-57},\; \alpha^{57})  &   (\alpha^{-59},\; \alpha^{5})  &   (\alpha^{-61},\; \alpha^{48})
 \end{array}$$}

Rational factorization returns three candidate rational functions.\\
(1). $\lambda(x)= 1+ \alpha^{15} x $ and $b(x)=\alpha^{31}+ \alpha^{24} x $.
The candidate error locator polynomial is constructed as
\begin{eqnarray*}
\Lambda'(x) \aline{=}  \lambda(x^2)\cdot \Lambda(x) + b(x^2)\cdot x^2B(x) \\
\aline{=} {1} + \alpha^{29} x  + \alpha^{59} x^{2} + \alpha^{38} x^{3} + \alpha^{2} x^{4} + \alpha^{46} x^{5} +
\alpha^{52} x^{6} + \alpha^{62} x^{7} + \alpha^{39} x^{8} + \alpha^{14} x^{9} + \alpha^{5} x^{10} \\
\aline{} + \alpha^{59} x^{11} + \alpha^{6} x^{12} + \alpha^{37} x^{13} \\
 \aline{=} (1-\alpha^{4}x)(1-\alpha^{7}x)(1-\alpha^{16}x)(1-\alpha^{17}x)(1-\alpha^{19}x)(1-\alpha^{20}x)(1-\alpha^{33}x)(1-\alpha^{45}x)(1-\alpha^{46}x) \\
\aline{}  (1-\alpha^{47}x)(1-\alpha^{50}x)(1-\alpha^{53}x)(1-\alpha^{60}x)
\end{eqnarray*}
which properly retrieves the prototype codeword.\\
(2). $\lambda(x) = 1+ \alpha^{52} x $ and $b(x)=\alpha^{55}$.
The candidate error locator polynomial is constructed as
\begin{eqnarray*}
\Lambda'(x) \aline{=} {1} + \alpha^{29} x  + \alpha^{29} x^{2} + \alpha^{7} x^{3} + \alpha^{8} x^{4} + \alpha^{11} x^{5} + a
^{11} x^{6} + \alpha^{15} x^{7} + \alpha^{15} x^{8} + \alpha^{28} x^{9} \\
\aline{} + \alpha^{17} x^{10} + \alpha^{13} x^{11} + \alpha^{23} x^{12} \\
\aline{=} (1-\alpha^{10}x)(1-\alpha^{12}x)(1-\alpha^{19}x)(1-\alpha^{24}x)(1-\alpha^{28}x)(1-\alpha^{31}x)(1-\alpha^{33}x)(1-\alpha^{36}x) \\
\aline{} (1-\alpha^{48}x)(1-\alpha^{49}x)(1-\alpha^{52}x)(1-\alpha^{59}x)  
\end{eqnarray*}
which yields an alternative candidate codeword with only 12-bit difference from the received word.
\begin{eqnarray*}
\bc \aline{=} [\; 1, 1, 0, 0, 1, 1, 0, 0, 0, 0, 0, 0, 0, 0, 1, 1, 1, 1, 1, 1, 1, 1, 1, 0, 0, 0, 0, 1, 1, 1, 1,  \\
	\aline{}  1, 1, 0, 0, 1, 1, 1, 1, 0, 0, 0, 1, 1, 0, 0, 0, 0, 1, 0, 1, 0, 1, 1, 0, 0, 0, 1, 1, 1, 0, 1, 0   \;].
\end{eqnarray*}
(3). $\lambda(x) = 1+ \alpha^{32} x + \alpha^{19} x^{2}$ and $b(x)=\alpha^{45}+ \alpha^{28} x $.
The corresponding candidate error locator polynomial is 
\begin{eqnarray*}
\Lambda'(x) \aline{=} {1} + \alpha^{29} x  + \alpha^{59} x^{2} + \alpha^{38} x^{3} + \alpha^{2} x^{4} + \alpha^{46} x^{5} +
\alpha^{52} x^{6} + \alpha^{62} x^{7} + \alpha^{39} x^{8} + \alpha^{14} x^{9} \\
\aline{} + \alpha^{5} x^{10} + \alpha^{59} x^{11} + \alpha^{6} x^{12} + \alpha^{37} x^{13}+\alpha^{53}x^{14}
\end{eqnarray*}
which does not contain precisely 14 distinct nonzero roots in $\GF(64)$ and thus is spurious.
 \hfill $\Box\Box$

\section{\sc Concluding Remarks}

Although the proposed list decoding algorithms are presented in the context of the list decoding of Reed-Solomon and BCH codes, 
their core is a rational curve-fitting algorithm, which may be viewed as
 an extension of the polynomial curve-fitting algorithm proposed in \cite{Guru-Sudan}.

Following the strategy of soft reliability transformation in the Koetter-Vardy algorithm \cite{Koetter-Vardy}, 
the proposed hard-decision list decoding algorithm can be  extended in a straightforward manner to 
algebraic soft-decision decoding using  weighted multiplicity array (in one-dimension), 
where weight is proportional to symbol error probability for (generalized) Reed-Solomon codes, 
or bit error probability for binary BCH codes. However, we may only obtain a one-dimensional multiplicity array 
due to the preprocessing of the Berlekamp-Massey algorithm, instead of a two-dimensional multiplicity matrix as in 
the Koetter-Vardy algorithm \cite{Koetter-Vardy}.

It is shown that the number of codewords is less than $n$ for list decoding up to the Johnson bound for binary codes \cite{Agrell, Elias, Guru-Sudan-bound},
whereas we have disclosed that the list size for decoding up to the Johnson bound is bounded by $O(t_\text{opt})$ 
for the particular class (nonbinary)  Reed-Solomon and binary BCH codes.  
We thus conjecture that $n$ is the universal bound for list decoding up to the Johnson bound for any code.   

Our developments are based on the notion of one-to-one correspondence along the two different definitions/interpretations of 
Reed-Solomon codes. Following the notion as well as the fact that BCH codes are subfield subcodes of Reed-Solomon codes \cite{Stich}, 
we conjecture that the Guruswami-Sudan algorithm can be modified in a way to achieve the Johnson bound for binary BCH codes.

\section*{Acknowledgements}

The author would like to thank Dr. Shih-Ming Shih, Prof. Paul Siegel, Prof. Ralf
Koetter, Prof. Jorn Justesen, Prof. Sergey Bezzateev, Prof. Ronny Roth, and specially Prof. Vladimir Sidorenko,  
for many constructive comments on improving the quality as well as the presentation of the manuscript.

\end{document}